\RequirePackage{fix-cm}
\documentclass[twocolumn,epjc3]{svjour3}
\let\oldbibitem\bibitem
\def\bibitem{\vfill\oldbibitem}
\RequirePackage{amsmath,amssymb}
\RequirePackage{hepnames}
\RequirePackage{textgreek}
\RequirePackage{upgreek}
\RequirePackage{graphicx}
\RequirePackage{mathptmx}
\RequirePackage{flushend}
\RequirePackage{hyperref}
\RequirePackage{siunitx}
\RequirePackage{booktabs}
\RequirePackage{multirow}
\RequirePackage{subfig}
\newcommand{\SIinterval}[3]{\ensuremath{[\num{#1};\ \num{#2}]\,\si{#3}}}
\newcommand{\itemlabel}[1]{\item\textbf{#1}\hspace{.5em}}
\newcommand{\figref}[1]{fig.~\ref{#1}}
\newcommand{\tabref}[1]{table~\ref{#1}}
\newcommand{\secref}[1]{section~\ref{#1}}
\renewcommand{\eqref}[1]{eq.~(\ref{#1})}
\journalname{Eur. Phys. J. C}
\DeclareMathOperator\const{const.}

\newcommand{\dx}[1]{\mathrm{d}{#1}}

\newcommand{\ie}{\mbox{i.\,e.}\xspace}

\newcommand{\bigO}[1]{\ensuremath{\mathcal{O}(#1)}}

\newcommand{\isotope}[2]{\texorpdfstring{\ensuremath{{}^{\text{#2}}\text{#1}}}{#1-#2}}
\newcommand{\Rn}[1]{\isotope{Rn}{#1}}

\newcommand{\Kr}[1]{\isotope{Kr}{#1}}
\newcommand{\Rb}[1]{\isotope{Rb}{#1}}

\newcommand{\tbeta}{\ensuremath{\upbeta}}

\newcommand{\CL}[1]{{#1}\%~\mbox{C.\,L.}}
\newcommand{\LN}[1]{\texorpdfstring{\ensuremath{\mathrm{LN}_{2}}}{LN2}}
\newcommand{\ExB}{\ensuremath{E \times B}}

\newcommand{\longit}[2][]{\ensuremath{{#2}^{\mathrm{#1}}_{\parallel}}}
\newcommand{\transv}[2][]{\ensuremath{{#2}^{\mathrm{#1}}_{\perp}}}
\newcommand{\magnit}[2][]{\ensuremath{|\vec{#2}^{\mathrm{#1}}|}}
\renewcommand{\slash}{\ensuremath{\,/\,}}

\newcommand{\MACE}{\mbox{MAC-E} filter\xspace}

\begin{document}
\title{Reduction of stored-particle background by a magnetic pulse method at the KATRIN experiment}
%
\thankstext{email}{\email{jan.behrens@kit.edu}}
%

\institute{%
Helmholtz-Institut f\"{u}r Strahlen- und Kernphysik, Rheinische Friedrich-Wilhelms Universit\"{a}t Bonn, Nussallee 14-16, 53115 Bonn, Germany\label{a}
\and Institute of Experimental Particle Physics~(ETP), Karlsruhe Institute of Technology~(KIT), Wolfgang-Gaede-Str. 1, 76131 Karlsruhe, Germany\label{b}
\and Institut f\"{u}r Kernphysik, Westf\"{a}lische Wilhelms-Universit\"{a}t M\"{u}nster, Wilhelm-Klemm-Str. 9, 48149 M\"{u}nster, Germany\label{c}
\and Institut f\"{u}r Physik, Johannes Gutenberg-Universit\"{a}t Mainz, 55099 Mainz, Germany\label{d}
\and Institute for Data Processing and Electronics~(IPE), Karlsruhe Institute of Technology~(KIT), Hermann-von-Helmholtz-Platz 1, 76344 Eggenstein-Leopoldshafen, Germany\label{e}
\and Institute for Nuclear Physics~(IKP), Karlsruhe Institute of Technology~(KIT), Hermann-von-Helmholtz-Platz 1, 76344 Eggenstein-Leopoldshafen, Germany\label{f}
\and Institute for Nuclear Research of Russian Academy of Sciences, 60th October Anniversary Prospect 7a, 117312 Moscow, Russia\label{g}
\and Institute for Technical Physics~(ITeP), Karlsruhe Institute of Technology~(KIT), Hermann-von-Helmholtz-Platz 1, 76344 Eggenstein-Leopoldshafen, Germany\label{h}
\and Max-Planck-Institut f\"{u}r Kernphysik, Saupfercheckweg 1, 69117 Heidelberg, Germany\label{i}
\and Max-Planck-Institut f\"{u}r Physik, F\"{o}hringer Ring 6, 80805 M\"{u}nchen, Germany\label{j}
\and Technische Universit\"{a}t M\"{u}nchen, James-Franck-Str. 1, 85748 Garching, Germany\label{k}
\and Laboratory for Nuclear Science, Massachusetts Institute of Technology, 77 Massachusetts Ave, Cambridge, MA 02139, USA\label{l}
\and Center for Experimental Nuclear Physics and Astrophysics, and Dept.~of Physics, University of Washington, Seattle, WA 98195, USA\label{m}
\and Nuclear Physics Institute of the CAS, v. v. i., CZ-250 68 \v{R}e\v{z}, Czech Republic\label{n}
\and Department of Physics, Faculty of Mathematics und Natural Sciences, University of Wuppertal, Gauss-Str. 20, D-42119 Wuppertal, Germany\label{o}
\and Department of Physics, Carnegie Mellon University, Pittsburgh, PA 15213, USA\label{p}
\and Universidad Complutense de Madrid, Instituto Pluridisciplinar, Paseo Juan XXIII, n\textsuperscript{\b{o}} 1, 28040 - Madrid, Spain\label{q}
\and Department of Physics and Astronomy, University of North Carolina, Chapel Hill, NC 27599, USA\label{r}
\and Triangle Universities Nuclear Laboratory, Durham, NC 27708, USA\label{s}
\and Commissariat \`{a} l'Energie Atomique et aux Energies Alternatives, Centre de Saclay, DRF/IRFU, 91191 Gif-sur-Yvette, France\label{t}
\and University of Applied Sciences~(HFD)~Fulda, Leipziger Str.~123, 36037 Fulda, Germany\label{u}
\and Department of Physics, Case Western Reserve University, Cleveland, OH 44106, USA\label{v}
\and Institute for Nuclear and Particle Astrophysics and Nuclear Science Division, Lawrence Berkeley National Laboratory, Berkeley, CA 94720, USA\label{w}
\and Institut f\"{u}r Physik, Humboldt-Universit\"{a}t zu Berlin, Newtonstr. 15, 12489 Berlin, Germany\label{x}
\and Project, Process, and Quality Management~(PPQ), Karlsruhe Institute of Technology~(KIT), Hermann-von-Helmholtz-Platz 1, 76344 Eggenstein-Leopoldshafen, Germany    \label{y}
}

\author{%
M.~Arenz\thanksref{a}
\and W.-J.~Baek\thanksref{b}
\and S.~Bauer\thanksref{c}
\and M.~Beck\thanksref{d}
\and A.~Beglarian\thanksref{e}
\and J.~Behrens\thanksref{email,c,f}
\and R.~Berendes\thanksref{c}
\and T.~Bergmann\thanksref{e}
\and A.~Berlev\thanksref{g}
\and U.~Besserer\thanksref{h}
% ...
}

\thankstext{also1}{Now at Wisconsin IceCube Particle Astrophysics Center~(WIPAC), 222 West Washington Ave., WI 53703, USA}
\thankstext{also2}{Also affiliated with Oak Ridge National Laboratory, Oak Ridge, TN 37831, USA}
%
\authorrunning{KATRIN Collaboration}
\date{Received: date / Accepted: date}
\maketitle
\begin{abstract}

The KATRIN experiment aims to determine the effective electron neutrino mass with a sensitivity of \SI{0.2}{eV/c^2} (\CL{90}) by precision measurement of the shape of the tritium \tbeta-spectrum in the endpoint region.
The energy analysis of the decay electrons is achieved by a \MACE spectrometer.
A common background source in this setup is the decay of short-lived isotopes, such as \Rn{219} and \Rn{220}, in the spectrometer volume.
Active and passive countermeasures have been implemented and tested at the KATRIN main spectrometer.
One of these is the magnetic pulse method, which employs the existing air coil system to reduce the magnetic guiding field in the spectrometer on a short timescale in order to remove low- and high-energy stored electrons.
Here we describe the working principle of this method and present results from commissioning measurements at the main spectrometer.
Simulations with the particle-tracking software \textsc{Kassiopeia} were carried out to gain a detailed understanding of the electron storage conditions and removal processes.

\keywords{%
Neutrino mass
\and Radon background
\and Background reduction methods
\and Monte Carlo methods
}
\end{abstract}
%
%
\section{Introduction}
\label{intro}

\begin{figure*}[tb]
    \centering
    \includegraphics[width=\textwidth]{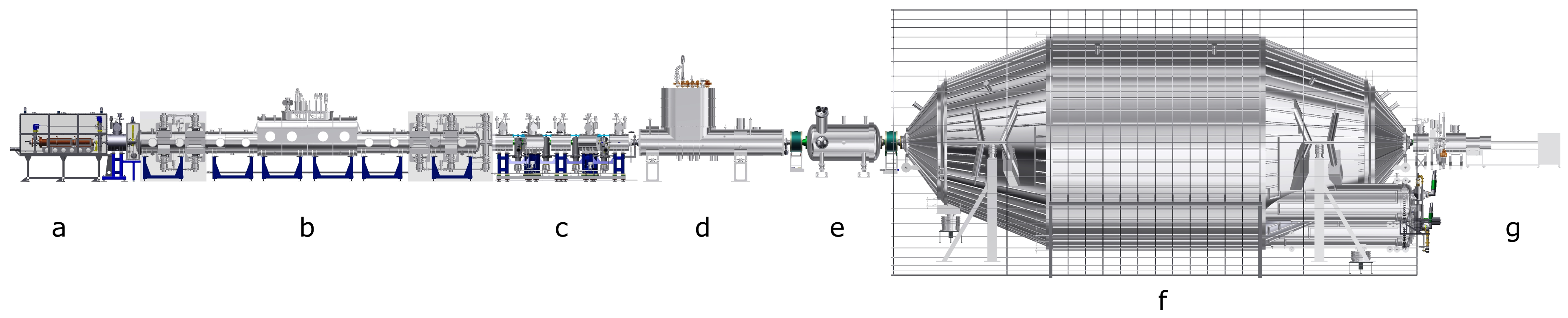}
    \caption{%
    The KATRIN beamline with a total length of about \SI{70}{m}: (a) rear section with calibration devices, (b) windowless gaseous tritium source, (c) differential pumping section, (d) cryogenic pumping section, (e) pre-spectrometer, (f) main spectrometer with air coil system, (g) focal-plane detector.
    The entire beamline transmits \tbeta-decay electrons in a \SI{191}{T.cm^2} flux tube to the detector.}
    \label{fig:katrin}
\end{figure*}

The \textbf{KA}rlsruhe \textbf{TR}itium \textbf{N}eutrino experiment \textbf{KATRIN}~\cite{KATRIN2005} aims to determine the `effective mass' of the electron neutrino (an incoherent sum over the mass eigenstates~\cite{Drexlin2013}) by performing kinematic measurements of tritium \tbeta-decay.
The target sensitivity of \SI{0.2}{eV/c^2} at \CL{90} improves the results of the Mainz~\cite{Kraus2005} and Troitsk~\cite{Aseev2011} experiments by one order of magnitude.

Figure~\ref{fig:katrin} shows the experimental setup of the KATRIN experiment.
Molecular tritium is injected into the windowless gaseous tritium source (WGTS~\cite{Priester2015,Heizmann2017}) where it undergoes \tbeta-decay.
The decay electrons emitted in the forward direction are adiabatically guided towards the spectrometer section in a magnetic field (\SI{191}{T.cm^2} flux tube) that is created by superconducting magnets.
The tritium flow into the spectrometer section is reduced by a factor of \num{e14}~\cite{Luo2006} by combining a differential pumping section (DPS~\cite{Lukic2012}) with a cryogenic pumping section (CPS~\cite{Gil2010,Roettele2017}).
A setup of \MACE\footnote{Magnetic Adiabatic Collimation with Electrostatic filter.} spectrometers~\cite{Picard1992,Lobashev1985} analyzes the kinetic energy of the decay electrons.
By combining an electrostatic retarding potential and a magnetic guiding field, the main spectrometer achieves an energy resolution of approximately \SI{1}{eV} at the tritium endpoint $E_0(\text{T}_2) = \SI{18574.00(7)}{eV}$~\cite{Otten2008,Myers2015}.
The magnetic field at the main spectrometer is achieved by superconducting magnets that are combined with air-cooled electromagnetic coils (air coils) surrounding the spectrometer vessel.
The high voltage of the main spectrometer is monitored by two precision high-voltage dividers that support voltages up to \SI{35}{kV} and \SI{65}{kV}, respectively~\cite{Thuemmler2009,Bauer2013a}.
The stability of the retarding potential is additionally monitored by another \MACE in a parallel beamline that measures \Kr{83m} conversion lines~\cite{Erhard2014}.

The integral \tbeta-spectrum is measured by varying the retarding potential near the tritium endpoint and counting transmitted electrons at the focal-plane detector (FPD~\cite{Amsbaugh2015}).
The FPD features a segmented layout with 148 pixels arranged in a dartboard pattern.
A post-acceleration electrode in front of the FPD allows to detect electrons with energies below its energy threshold of about \SI{7}{keV}.
The neutrino mass is determined by fitting the convolution of the theoretical \tbeta-spectrum with the response function of the entire apparatus to the data~\cite{PhDKleesiek2014}.
This takes into account parameters such as the final states distribution of $\text{T}_2$ decay, the energy loss spectrum of the WGTS, and other systematic corrections~\cite{KATRIN2005,Babutzka2012,SSCPaper}.

An intrinsic disadvantage of the \MACE setup is a high storage probability of high-energy electrons that are created in the spectrometer volume from, for example, nuclear decays~\cite{Mertens2013}.
During their long storage times of up to several hours~\cite{PhDWandkowsky2013}, these electrons can create low-energy secondaries via scattering processes with residual gas.
The retarding potential accelerates these electrons towards the detector, where they reach a kinetic energy close to the tritium endpoint.
This background is indistinguishable from signal \tbeta-electrons, and dedicated countermeasures are needed to reach the KATRIN sensitivity goal.
In addition to passive background reduction techniques such as \LN{}-cooled baffles~\cite{Arenz2016}, active methods that remove stored electrons from the flux tube volume have been implemented at the main spectrometer.
The ``magnetic pulse method'' aims to break the storage conditions of low- and high-energy electrons, which results in their removal from the spectrometer volume.
Because the method interferes with the process of the \tbeta-spectrum measurement, it should be applied in a reasonably short timescale on the order of a few seconds.
This is achieved by utilizing the existing air coil system~\cite{Erhard2018} to invert the magnetic guiding field, which forces electrons towards the vessel walls where they are subsequently captured.
The current inversion is performed by an electronic current-inverter device for air coil currents up to \SI{180}{A} that was developed at WWU M{\"u}nster and KIT~\cite{PhDBehrens2016}.

In this article we discuss the technical design of the magnetic pulse system at the main spectrometer and its integration into the existing air coil setup (\secref{design}).
We present measurement results from two commissioning phases of the KATRIN spectrometer section, where we investigated the removal efficiency with artificially enhanced background (using \Kr{83m} and \Rn{220} sources) and the background reduction under nominal conditions (\secref{measurements}).
To further investigate the effects of the magnetic pulse on stored electrons, we discuss simulations with the particle-tracking software \textsc{Kassiopeia}, which has been developed by the KATRIN collaboration~\cite{Furse2017} (\secref{simulations}).
%
\section{Setup and design}
\label{design}

\begin{figure*}[t]
    \centering
    \includegraphics[width=\columnwidth]{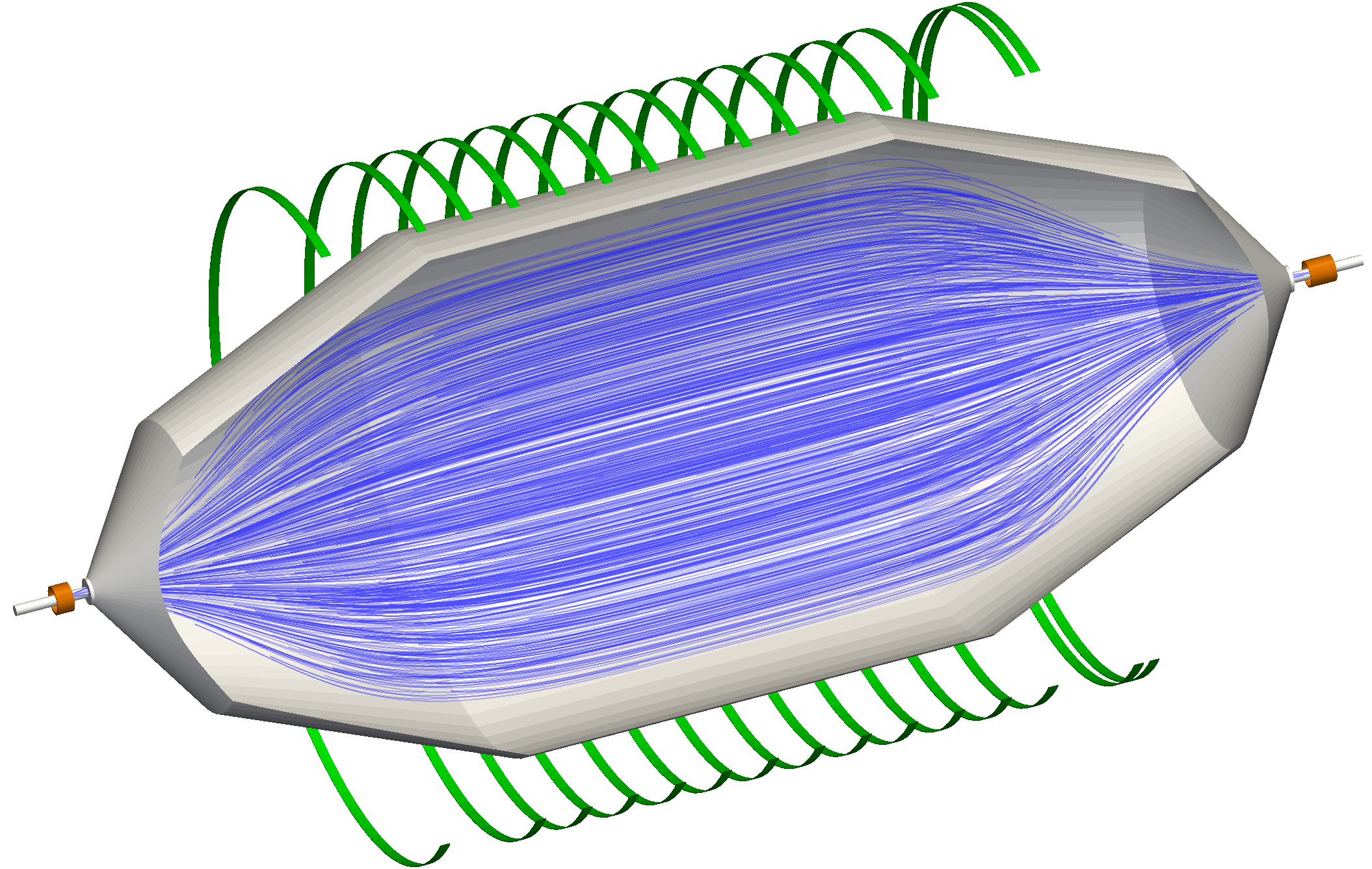}
    \includegraphics[width=\columnwidth]{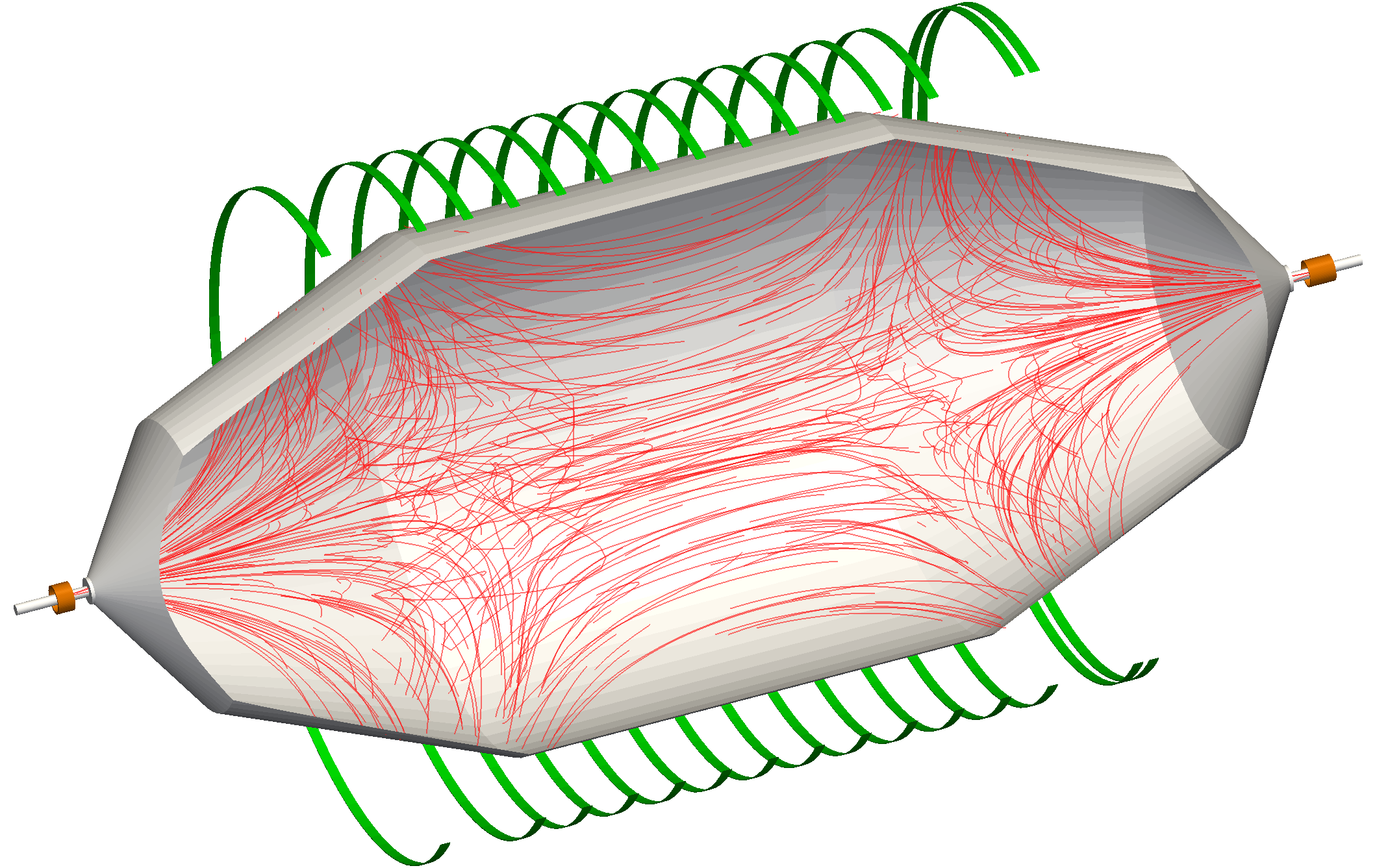}
    \caption{Deformation of the flux tube by the magnetic pulse method.
        \emph{Left:} Under nominal conditions, the \SI{191}{T.cm^2} flux tube (blue field lines) is fully contained inside the main spectrometer vessel.
        \emph{Right:} The flux tube is deformed when the magnetic field is reduced by the magnetic pulse.
            This is achieved by inverting the currents of the LFCS air coils (green) while keeping the superconducting solenoids (orange) at either end of the spectrometer at nominal field.
        The field inversion causes the magnetic field lines (red) to connect to the vessel walls, which removes stored electrons from the spectrometer volume.}
    \label{fig:magpulse_fluxtube}
\end{figure*}

\subsection{Background from stored particles}
\label{design:background}

The sensitivity of KATRIN is significantly constrained by background processes in the spectrometer section, which contribute to the statistical uncertainty of the determined neutrino mass~\cite{PhDKleesiek2014}.
Passive and active methods to reduce this background component have been implemented at the main spectrometer and were investigated during several commissioning measurement phases~\cite{PhDBehrens2016,PhDGoerhardt2014,PhDHarms2015,PhDHilk2016}.
Earlier investigations have shown that a major background component arises from $\alpha$-decays of radon isotopes \Rn{219} and \Rn{220} inside the spectrometer volume.
Each decay can release electrons by various processes: \emph{conversion electrons} with energies up to \SI{450}{keV}, \emph{shake-off electrons} with energies up to \SI{80}{keV}, \emph{Auger electrons} with energies up to \SI{20}{keV}, and \emph{shake-up electrons}  with energies up to \SI{230}{eV} that are emitted due to reorganization of atomic electrons.
By these processes radon decays typically produce high-energy primary electrons with energies up to several hundred \si{keV}.
These can in turn produce low-energy secondary electrons via scattering processes with residual gas~\cite{Wandkowsky2013}.
At the main spectrometer, the residual gas composition is dominated by hydrogen~\cite{Arenz2016}.

The main advantage of the \MACE is its excellent energy resolution of $\Delta E \approx \SI{1}{eV}$ at the tritium endpoint $E_0(\mathrm{T_2})$, as given by the magnetic field ratio
\begin{equation}
    \label{eq:energy_resolution}
    \frac{\Delta E}{E} = \frac{B_\mathrm{min}}{B_\mathrm{max}}
    \,.
\end{equation}
Here $E \approx E_0(\mathrm{T_2})$ is the kinetic energy of the signal electrons, $B_\mathrm{min}$ is the magnetic field at the central plane of the main spectrometer -- the analyzing plane -- where the energy analysis occurs, and $B_\mathrm{max}$ is the maximum magnetic field in the beam line (at the pinch magnet located at the spectrometer exit).
Under nominal conditions, the field strengths are $B_\mathrm{min} = \SI{0.3}{mT}$ and $B_\mathrm{max} = \SI{6}{T}$.

Unfortunately, the \MACE also provides highly favorable storage conditions for electrons that are created inside the spectrometer volume.
The adiabatic collimation, which is a key feature of the energy analysis, can cause a magnetic reflection of electrons at the spectrometer entrance and exit.
This process is known as the ``magnetic bottle'' effect.
It is best described in terms of the electron pitch angle $\theta = \angle(\vec{p},\vec{B})$, which relates to its longitudinal and transverse kinetic energy $\longit{E},\ \transv{E}$:
\begin{equation}
    \label{eq:long_trans_energy}
    \longit{E}  = E \cdot \cos^2 \theta
    \,,
    \quad
    \transv{E}  = E \cdot \sin^2 \theta
    \,,
\end{equation}
where $E$ denotes the electron's kinetic energy, $\vec{p}$ its momentum, and $\vec{B}$ is the local magnetic field.
Electrons are stored in the \MACE if if $\transv{E} > \Delta E$.

To describe the \MACE one can define an adiabatic constant that is conserved during propagation,
\begin{equation}
    \label{eq:magnetic_moment}
    \gamma \mu  = \frac{\gamma + 1}{2} \cdot \frac{\transv{E}}{\magnit{B}} = \const
    \,,
\end{equation}
where $\mu$ denotes the magnetic moment of a gyrating electron~\cite{Picard1992}.
This results in a pitch angle increase when the electron moves into a higher magnetic field.
For $\si{keV}$ electrons, the relativistic gamma factor is $\gamma \lesssim \num{1.04}$, so the non-relativistic approximation can be applied.

Electrons change their direction of propagation if their pitch angle reaches \ang{90}.
For electrons created in the spectrometer at a magnetic field $B_0$ with a pitch angle $\theta_0$, this occurs if $\theta_0 \ge \theta_\mathrm{max}$ with
\begin{equation}
    \label{eq:magnetic_bottle}
    \theta_\mathrm{max} = \arcsin\sqrt{ \frac{B_0}{B_\mathrm{mag}} }
\end{equation}
in adiabatic approximation, where $B_\mathrm{mag}$ is the magnetic field at the spectrometer entrance or exit\footnote{%
    In the KATRIN setup, the nominal magnetic field at the entrance (\SI{4.5}{T}) is smaller than at the exit ($B_\mathrm{max} = \SI{6.0}{T}$).
    Electrons from the spectrometer therefore have a higher escape probability at the entrance.}.
Electrons from nuclear decays follow an isotropic emission profile, and thus have large pitch angles on average.
They are stored with high efficiency by the magnetic bottle effect.

The relation in \eqref{eq:magnetic_bottle} only applies if the high-field region is at ground potential ($U = 0$), which is the case at the main spectrometer.
Electrons originating from the spectrometer volume are accelerated by the retarding potential ($U_0 \approx \SI{-18.6}{kV}$) towards the grounded beamline.
Secondary electrons  with small initial kinetic energies that arrive at the FPD appear in the energy interval near the tritium endpoint, together with signal electrons from tritium \tbeta-decay.
This indiscriminable background follows a non-Poissonian distribution, which significantly enhances its impact on the neutrino mass sensitivity.
It is vital to suppress this background contribution in order to achieve the desired KATRIN sensitivity~\cite{Mertens2013}.


\subsection{Background reduction methods}
\label{design:bgreduction}

\begin{figure*}[t]
    \centering
    \includegraphics[width=\columnwidth]{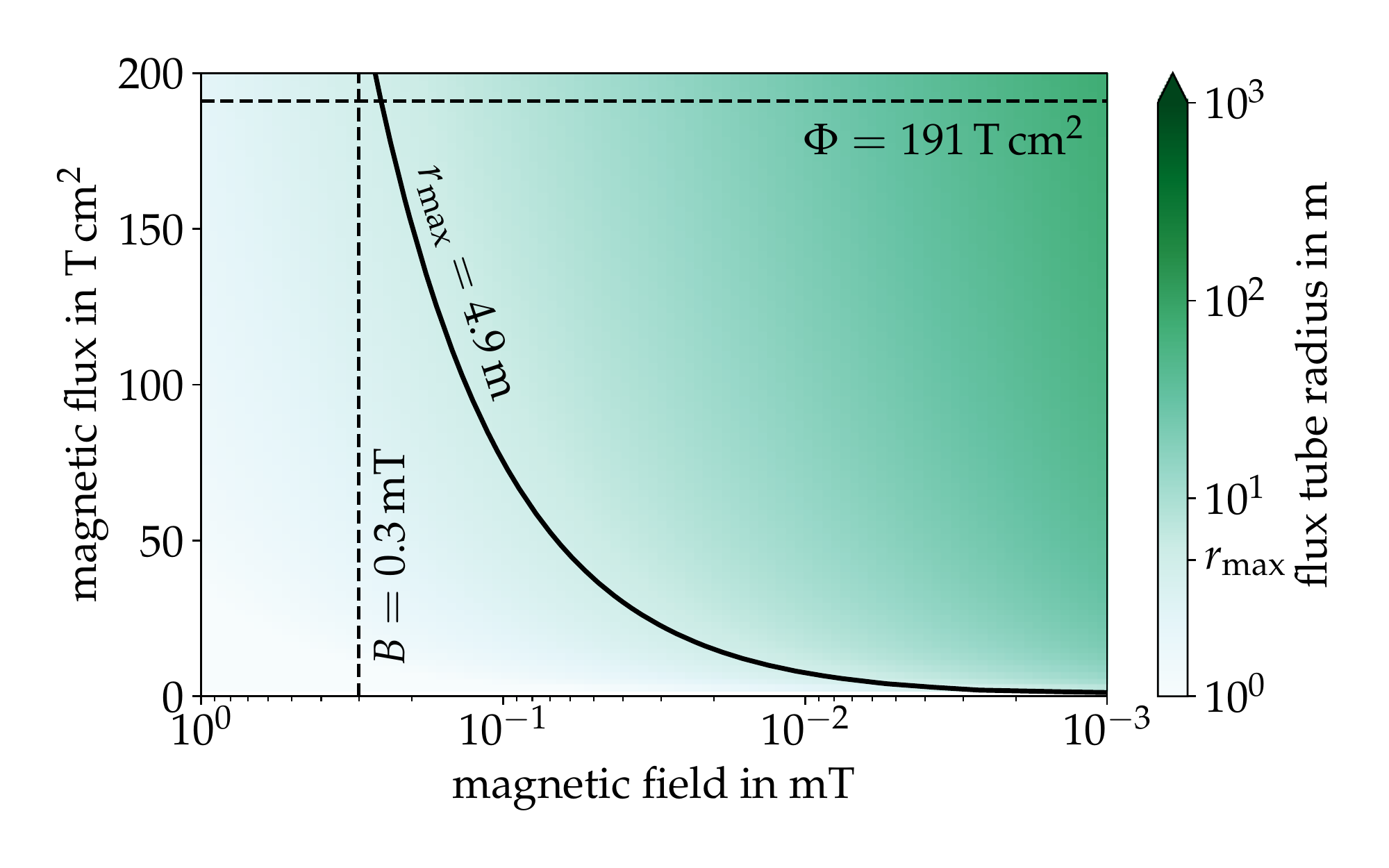}
    \includegraphics[width=\columnwidth]{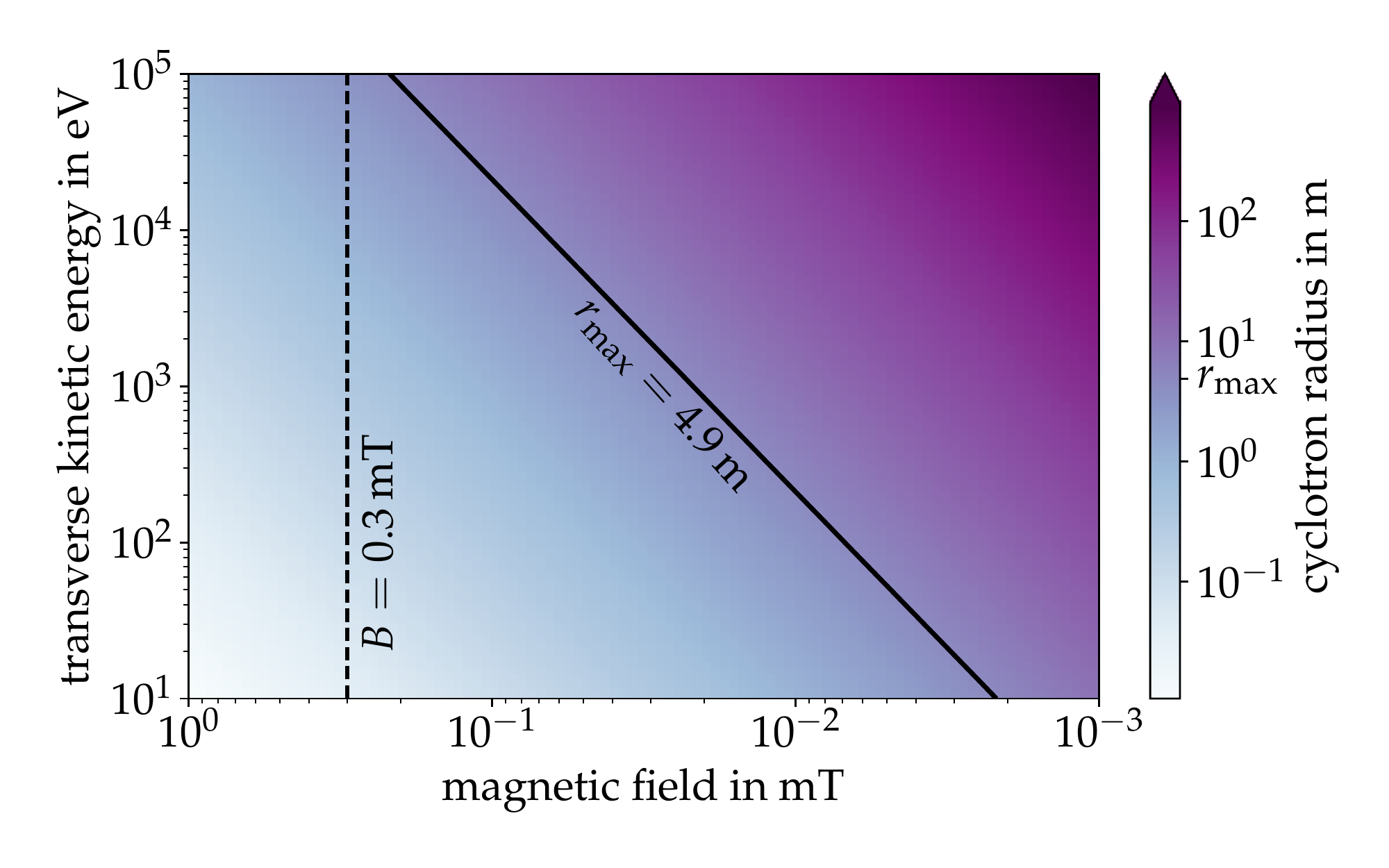}
    \caption{Effects of the magnetic pulse on stored electrons.
        The magnetic field decreases towards the right in both plots; the dashed vertical lines mark the nominal field in the analyzing plane ($B_\mathrm{min} = \SI{0.3}{mT}$).
        \emph{Left:} The flux tube widens when the magnetic field decreases according to \eqref{eq:flux_tube}.
            In this plot the magnetic flux corresponds to a flux tube cross-section with radius $r_\mathrm{flux}$ in the spectrometer center (indicated by color), and the solid line indicates the maximal vessel radius $r_\mathrm{max} = \SI{4.9}{m}$.
            Electrons stored in a flux tube region with $r_\mathrm{flux} > r_\mathrm{max}$ are removed, and electrons stored at smaller radii require a larger field reduction to be removed.
            The dashed horizontal line marks the conserved flux of \SI{191}{T.cm^2} under nominal conditions.
        \emph{Right:} The cyclotron radius \eqref{eq:cyclotron_radius} increases when the magnetic field is reduced.
            The solid line again corresponds to the vessel radius and indicates the removal threshold.
            This effect depends on the transverse kinetic energy of the electrons.
            A stronger magnetic field reduction is necessary to remove low-energy electrons via this process.}
    \label{fig:magpulse_theory}
\end{figure*}

A major source of radon nuclei in the main spectrometer is the array of non-evaporable getter (NEG) pumps, which are used in combination with turbo-molecular pumps (TMPs) to achieve ultra-high vacuum conditions down to $p \le \SI{e-10}{mbar}$.
Small amounts of radon are also released from welds in the vessel walls.
The short half-life of \Rn{219} ($t_{1/2} = \SI{3.96}{s}$) and \Rn{220} ($t_{1/2} = \SI{55.6}{s}$) allows these nuclei to enter and decay inside the spectrometer volume before being pumped out by the vacuum system, which achieves a typical turn-around time of \SI{350}{s}~\cite{PhDBehrens2016}.
To reduce the background that originates from radon decays, \LN{}-cooled baffles are mounted in front of the NEG pump ports as a passive countermeasure.
It was found that the baffles block the majority of radon nuclei from entering the spectrometer, establishing a reduction of radon-induced background by about \SI{95}{\%}~\cite{Drexlin2017}.

As additional countermeasures against stored-particle background, two active background removal methods have been implemented at the main spectrometer.
They provide a complementary approach to remove stored electrons from the spectrometer volume:
\begin{itemize}
    \itemlabel{Electric dipole method}
        A dipole field is applied inside the spectrometer volume, using the inner-electrode (IE) system that is mounted on the inner vessel walls~\cite{Valerius2010}.
        It is split into half-ring segments where a voltage difference of up to \SI{1}{kV} can be applied. The resulting dipole field $E_\mathrm{dip} \le \SI{100}{V/m}$ in combination with the magnetic guiding field induces an \ExB{} drift.
        This method is efficient at removing low-energy electrons~\cite{PhDHilk2016,PhDWandkowsky2013}.

    \itemlabel{Magnetic pulse method}
        The magnetic field inside the spectrometer volume is reduced via the low-field correction system (LFCS, 14 circular air coils) or the earth magnetic field compensation system (EMCS, 2 wire loops) that enclose the spectrometer vessel~\cite{Erhard2018}.
        By inverting the electric current in the air coils it is possible to reduce the magnetic guiding field in the spectrometer central section on short time scales of about \SI{1}{s}.
        The effects on stored electrons are discussed below.
        This method efficiently removes high- and low-energy electrons.
\end{itemize}

Figure~\ref{fig:magpulse_fluxtube} illustrates the deformation of the flux tube by the magnetic pulse method.
Under nominal conditions, the magnetic flux tube is fully contained inside the spectrometer vessel (left panel).
Reducing or inverting the magnetic guiding field deforms the flux tube (right panel).
The field reduction by the magnetic pulse causes the following three effects that can lead to the removal of stored electrons from the spectrometer:
\begin{enumerate}

    \itemlabel{Flux tube size}
        The magnetic flux of $\Phi = \SI{191}{T.cm^2}$ is conserved over the entire beam line of the experiment,
        \begin{equation}
            \label{eq:flux_tube}
            \Phi = \oint \vec{B} \,\dx{\vec{A}} \approx B \cdot \pi r_\mathrm{flux}^2 = \const
            \,,
        \end{equation}
        where $A = \pi r_\mathrm{flux}^2$ is the cross-section of the flux tube with radius $r_\mathrm{flux}$ at a given magnetic field $B$.
        The approximation assumes that the magnetic field is homogeneous over the entire cross-section.
        The flux tube is contained inside the spectrometer under nominal conditions ($r_\mathrm{flux} \le r_\mathrm{max}$ at $B = B_\mathrm{min}$, where $r_\mathrm{max} = \SI{4.9}{m}$ is the radius of the spectrometer vessel in the analyzing plane).
        A decrease in the magnetic field results in a widening of the flux tube, so that the outer magnetic field lines connect to the vessel walls.
        This removes electrons that are stored in the outer regions of the flux tube at high radii; electrons stored at smaller radii require a larger field reduction.
        Hence, this effect features a radial dependency, and the overall removal efficiency increases as the magnetic field is reduced (left panel of \figref{fig:magpulse_theory}).

        This process is very efficient for the removal of stored electrons, and is in fact the dominant effect of the magnetic pulse method.
        However, the removal efficiency is limited because (even for a fully inverted field, $B_\mathrm{min} < 0$) electrons can be magnetically reflected according to \eqref{eq:magnetic_bottle} before they reach the vessel walls.
        Hence, a fraction of stored electrons at small radii typically remain inside the flux tube.
        This limitation was investigated by particle-tracking simulations and is further discussed in \secref{simulations}.

    \itemlabel{Cyclotron motion}
        An electron moving in a magnetic guiding field undergoes a cyclotron motion (gyration) around a magnetic field line.
        The cyclotron radius $r_c$ is defined in approximation as
        \begin{equation}
            \label{eq:cyclotron_radius}
            r_c = \dfrac{\transv{p}}{|q| \, |B|}
            \,,
        \end{equation}
        where $\transv{p}$ and $q$ are the transverse momentum and the charge of the electron, and $|B|$ is the magnitude of the local magnetic field.
        The transverse momentum depends on the electron's kinetic energy ($\transv{p} \approx \sqrt{2 m \transv{E}}$).
        The cyclotron radius increases when the magnetic field is reduced, so that electrons are removed if $r_c \gtrsim r_\mathrm{max}$.

        This effect depends on kinetic energy (and also pitch angle) of the electron and is more efficient for high-energy electrons (right panel of \figref{fig:magpulse_theory}).
        It has a radial dependence since electrons stored at higher radii are closer to the vessel walls, hence a smaller field reduction is needed to remove these electrons.

    \itemlabel{Induced radial drift}
        According to the third Maxwell equation, $\nabla \times \vec{E}_\mathrm{ind} = -\dot{\vec{B}}$,
        a magnetic field change $\dot{\vec{B}}$ induces an electric field $\vec{E}_\mathrm{ind}$.
        At the main spectrometer the induced electric field is mainly oriented in azimuthal direction, $|\vec{E_\mathrm{ind}}| \approx E_\phi$, since $\vec{B} \approx (0, 0, B_z)$ in the spectrometer central region:
        \begin{equation}
            \label{eq:magpulse_induced_field}
            E_\mathrm{ind} \approx E_\phi = -\frac{r}{2} \cdot \dot{B}_z
            \,,
        \end{equation}
        where $r = \sqrt{x^2 + y^2}$ denotes the radial position of an electron in the magnetic field and corresponds to its distance from the spectrometer symmetry axis.
        The combination of the induced electric and the magnetic guiding field causes a drift of electrons in the spectrometer,
        \begin{equation}
            \vec{v}_\mathrm{drift} = \dfrac{\vec{E_\mathrm{ind}} \times \vec{B}}{B^2}
        \end{equation}
        in adiabatic approximation.
        It is mainly oriented in the radial direction due to the combination of axial magnetic field and azimuthal electric field from \eqref{eq:magpulse_induced_field}:
        \begin{equation}
            \label{eq:magpulse_drift}
            v_\mathrm{drift} \approx \dfrac{E_\phi \cdot B_z}{B_z^2} = -\dfrac{r}{2} \cdot \dfrac{\dot{B}_z}{B_z}
            \,.
        \end{equation}
        The drift is directed outwards in a reducing magnetic field (during the magnetic pulse) and directed inwards in an increasing magnetic field (after the magnetic pulse, when the field returns to nominal).
        Hence, electrons move towards the vessel walls while a magnetic pulse is applied, which contributes to the overall removal efficiency.

        This removal process is independent of the electron energy, but is more efficient for electrons stored on outer field lines at larger radii.
        The drift speed is highly time-dependent because of the exponential behavior of $B(t)$ (\secref{simulations:setup}) and maximal only for a short time where $|B| \rightarrow 0$ during field inversion.
        In comparison with the other two discussed effects, the induced drift only plays a minor role in the removal of stored electrons.
\end{enumerate}

\subsection{The magnetic pulse system}
\label{design:magpulse}

To apply a fast magnetic field change at the main spectrometer, the magnetic pulse method utilizes the existing air coil system.
The air coil system was implemented to allow fine-tuning of the spectrometer's transmission properties by varying individual air coil currents~\cite{Glueck2013}.
The LFCS permits one to vary the nominal magnetic field $B_\mathrm{min}$ in the analyzing plane in a range of \SIrange{0}{2}{mT}.
The air coils are operated at currents of up to \SI{175}{A}, which are generated by individual power supplies.

To implement the magnetic pulse method, the air coil system was extended to allow a fast inversion of the magnetic guiding field~\cite{PhDBehrens2016}.
This is achieved by inverting the current direction in the air coils, employing dedicated current-inverter units (``flip-boxes'') for each air coil that allow a fast switching of the coil current direction without changing the absolute current.
A detailed description of the system has been published in \cite{Erhard2018}.
The independent units are installed between each air coil and its corresponding power supply, and have been integrated with the air coil slow-control system for remote operation.
The precise timestamps of the trigger signals to each individual flip-box are fed into the DAQ system and stored with the measurement data.
The timing information is thus readily available for subsequent data analysis.
%
\section{Measurements}
\label{measurements}

The magnetic pulse method that we present here has been tested successfully in two commissioning phases of the KATRIN spectrometer section.
The measurements in phase~I were performed in 2013 with a single flip-box prototype.
This allowed us to perform functionality tests and to investigate the magnetic field reduction in a preliminary setup.
A radioactive \Kr{83m} source was mounted at the spectrometer to artificially increase the background rate for these measurements.
In the phase~II, carried out in 2014/2015, the complete magnetic pulse system with all air coils equipped with flip-boxes was available.
Here we further investigated the removal efficiency of the magnetic pulse with a \Rn{220} source at the spectrometer.
We also examined the reduction of the nominal spectrometer background and performed measurements in combination with an electron source (see \cite{Behrens2017} for a description of the device), where we investigated the magnetic pulse timing inside the spectrometer vessel with an electron beam~\cite{PhDBehrens2016}.

\subsection{Magnetic fields at the spectrometer}
\label{measurements:test}

A magnetic field measurement was performed with a fast fluxgate sensor (Bartington Instruments Mag-03 three-axis sensor, \SI{1}{ms} sampling interval) during preparation for the phase~II measurements.
The intention was to verify the functionality of the magnetic pulse system and to determine the time constant of the magnetic field inversion.
The sensor was mounted on the outer wall of the spectrometer vessel ($r \approx r_\mathrm{max} = \SI{4.9}{m}$) close to the analyzing plane; details are given in \cite{Erhard2018}.
Under nominal conditions, the magnetic field at the sensor position was $B_0 = \SI{0.36}{mT}$.
To apply a magnetic pulse, the LFCS air coils L1--L13 were inverted simultaneously by the flip-box units\footnote{%
    The air coil L14, which is closest to the detector, is already inverted under nominal conditions to compensate the strong magnetic field of the pinch magnet.}.
The measurement shows that the magnetic field change can be described by a model that is the sum of two exponential decay curves with a fast time constant $\tau_\mathrm{fast} = \SI{29.6(1)}{ms}$ and a slow time constant $\tau_\mathrm{slow} = \SI{419.8(4)}{ms}$.
This model is also employed in the particle-tracking simulations and discussed in \secref{simulations:setup}.

The direct field measurement shows that on the outside of the spectrometer, magnetic field inversion ($B \approx 0$) is achieved within $< \SI{100}{ms}$ after initiating the magnetic pulse.
After a relaxation time of about \SI{1}{s}, the magnetic field reaches $B < \SI{-0.29}{mT}$ and remains at the inverted level, until the magnetic pulse ends and the field goes back to its nominal strength.
The total pulse amplitude of \SI{0.65}{mT} at the sensor position is limited by the stray field of the super-conducting solenoids at the spectrometer and by the maximum air coil currents.
Because a significant reduction of the magnetic field is required to remove stored electrons, the typical timescale of the magnetic pulse is \SIrange{500}{1000}{ms}.
Hence, the effective behavior of the magnetic field for $t > \SI{100}{ms}$ is fully described by the slow time constant.
It is expected that the magnetic field \emph{inside} the spectrometer, which cannot be measured directly, shows a very similar behavior with larger time constants due to additional effects such as eddy currents in the vessel hull.
This is discussed in \secref{measurements:egun}.

\subsection{Phase I: Measurements with a radioactive \Kr{83m} source}
\label{measurements:krypton}

In the first commissioning phase of the main spectrometer, the removal efficiency of the magnetic pulse was investigated by attaching a \Rb{83} emanator~\cite{Sentkerestiova2018} at one of the pump ports to increase the background artificially.
The emanator produces radioactive \Kr{83m} nuclei ($t_{1/2} = \SI{1.8}{h}$) that propagate into the spectrometer volume.
The subsequent nuclear decays produce stored-electron background that is similar to the background expected from radon decays: high-energy primary electrons are produced as conversion electrons predominantly by three lines: $E_\mathrm{L_1-9.4} = \SI{7.73}{keV}$, $E_{L_3-32} = \SI{30.48}{keV}$, and $E_\mathrm{K-32} = \SI{17.82}{keV}$~\cite{Venos2018}.

Scattering processes with residual gas create additional low-energy secondary electrons that become stored inside the spectrometer due to the magnetic bottle effect (\secref{design}).
A low fraction of electrons leave the spectrometer towards the detector system where they are observed (see \secref{simulations:efficieny}).
Their energy spectrum at the detector is shifted by the retarding potential ($U_\mathrm{ana} = \SI{-18.6}{kV}$), the post-acceleration voltage ($U_\mathrm{PAE} = \SI{10}{kV}$), and the detector bias voltage ($U_\mathrm{FPD} = \SI{120}{V}$); this yields an energy shift of $q (U_\mathrm{ana} - U_\mathrm{PAE} - U_\mathrm{FPD}) = \SI{28.72}{keV}$.
Hence, secondary electrons with $E \approx \SI{0}{eV}$ are observed at a peak energy of \SI{28.72}{keV}, and the same shift applies to high-energy \Kr{83m} primary electrons.
Table~\ref{tab:measurements:roi} lists the corresponding region of interest (ROI) that was applied as an energy selection for primary and secondary electrons.

\begin{table}[h]
    \caption{Region of interest (ROI) for low-energy secondary electrons and high-energy primary electrons from \Kr{83m} decays, applied as an energy selection $[E_\mathrm{min};\ E_\mathrm{max}]$ to the detector data. The high-energy range is combined into a single ROI.}
    \newcolumntype{L}[1]{>{\raggedright\arraybackslash}p{#1}}
    \newcolumntype{C}[1]{>{\centering\arraybackslash}p{#1}}
    \newcolumntype{R}[1]{>{\raggedleft\arraybackslash}p{#1}}
    \newcommand{\ctab}{\centering\arraybackslash}
    \centering
    \begin{tabular}{lrrrr}
    \toprule
         \ctab $E_0\slash\si{keV}$ 
        &\ctab $E_\mathrm{det}\slash\si{keV}$
        &\ctab $E_\mathrm{min}\slash\si{keV}$
        &\ctab $E_\mathrm{max}\slash\si{keV}$ \\
    \midrule
        secondary $\Pelectron$  &\num{0}        &\num{28.72}    &\num{25.72}    &\num{30.72}    \\
    \midrule
        $\mathrm{L_1\!-\!9.4}$  &\num{7.48}     &\num{36.45}    & $\quad|\quad$ & $\quad|\quad$ \\
        $\mathrm{K\!-\!32}$     &\num{17.82}    &\num{46.54}    &\num{33.45}    &\num{61.20}    \\
        $\mathrm{L_3\!-\!32}$   &\num{30.48}    &\num{59.20}    & $\quad|\quad$ & $\quad|\quad$ \\
    \bottomrule
    \end{tabular}
    \label{tab:measurements:roi}
\end{table}

For these measurements only one flip-box prototype was available.
The removal efficiency of the magnetic pulse is limited since only one LFCS air coil (L8) could be inverted with this setup.
The measurement allowed us to study the effect on stored electrons in different energy regimes.
It is especially interesting to compare the removal of low- and high-energy electrons.
When examining the background rate observed at the detector, one must consider that a rate reduction does not necessarily imply a removal of electrons from the spectrometer volume, since only the non-stored electrons that escape to the detector can be observed.
A reduction in background rate could also be explained by modified storage conditions while the magnetic field is reduced.
Hence, the actual amount of removed electrons must be determined by comparing the observed rate before and after a magnetic pulse when the magnetic field is undisturbed.
Because the electromagnetic conditions are the same in these cases, the electron rates can be compared directly and an observed rate reduction can be safely attributed to the removal of stored electrons from the flux tube.

\begin{figure}[t]
    \centering
    \includegraphics[width=\columnwidth]{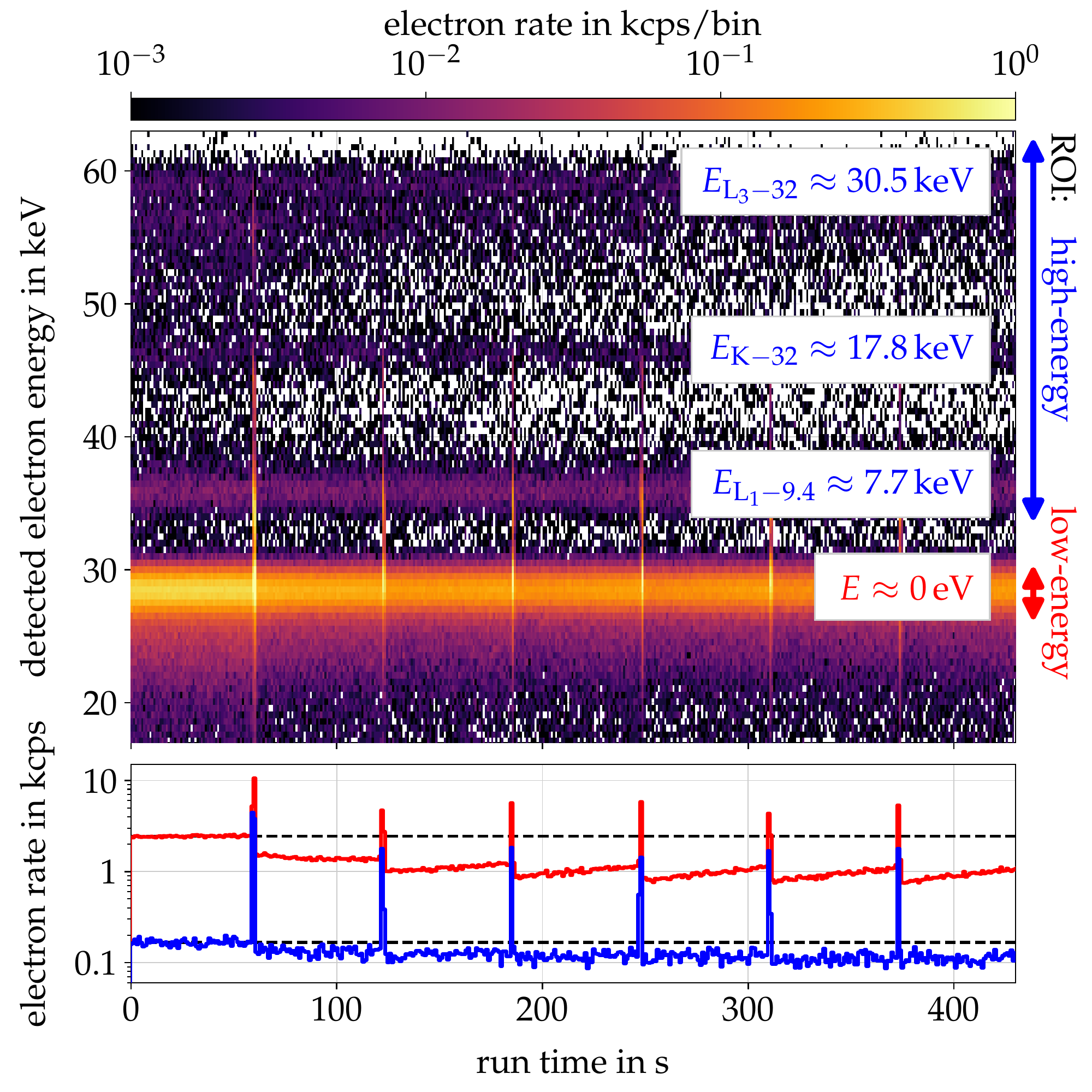}
    \caption{Removal of stored electrons originating from a \Kr{83m} source by the magnetic pulse.
        The pulse was applied by inverting LFCS air coil L8 for \SI{1}{s} with the flip-box prototype.
        \emph{Top}: The energy spectrum shows lines of high-energy primary electrons from \Kr{83m} decay ($E = \SIinterval{7.48}{32.14}{keV}$) and low-energy secondary electrons ($E \approx \SI{0}{eV}$).
        \emph{Bottom:} The observed rate in both energy regimes -- low-energy (red) and high-energy (blue) -- is reduced by applications of the magnetic pulse.
        The rate increases after each pulse from continuous \Kr{83m} decays in the spectrometer.
        The ``rate spikes`` are a result of the flux tube deformation.}
    \label{fig:measurements:sds1_krypton}
\end{figure}

The top panel of \figref{fig:measurements:sds1_krypton} shows the observed energy spectrum over time while several magnetic pulses are applied (one pulse of length \SI{1}{s} every $\sim \SI{60}{s}$).
The spectrum shows distinct lines of low-energy secondary and high-energy primary electrons from \Kr{83m} decay.
The bottom panel shows the total electron rate in the low- and high-energy ROI.
In this setting, the nominal background rate before the application of magnetic pulses is $\dot{N}_0 = \SI{2.43(5)}{cps}$ in the low-energy regime and $\dot{N}'_0 = \SI{0.17(1)}{cps}$ in the high-energy regime.
Each application of the magnetic pulse results in a rate reduction due to the removal of stored electrons; this is especially visible at low energies.
Continuous nuclear decays in the spectrometer then result in a gradual rate increase after each pulse.
Additionally, a pronounced ``rate spike'' is observed during a pulse, which is caused by the flux tube deformation that allows electrons from the vessel walls to reach the detector directly\footnote{%
    These electrons, arising from mechanisms such as cosmic muons hitting the vessel walls, are blocked from entering the inner spectrometer volume by the magnetic guiding field under nominal conditions.}.
The rate spikes are therefore a useful indicator of the functionality of the magnetic pulse system.

The difference in the observed rate before ($\dot{N}_0$) and after the pulse ($\dot{N}_\mathrm{min}$) allows the determination of the removal efficiency, defined here as the ratio
\begin{equation}
    \label{eq:removal_efficiceny}
    R = \frac{\dot{N}_0 - \dot{N}_\mathrm{min}}{\dot{N}_0}
    \,.
\end{equation}
The rates are determined by averaging the observed rate over \SI{5}{s} before and after each application of the magnetic pulse; results are shown in \tabref{tab:measurements:sds1_krypton} for the low-energy regime.
The repetition interval of \SI{60}{s} is shorter than the relaxation time required to reach the nominal rate $\dot{N}_0$ after each magnetic pulse, and the absolute number of stored electrons decreases in subsequent pulse cycles as shown in fig.~\ref{fig:measurements:sds1_krypton}.
After three pulse cycles the observed rate follows a repetitive pattern, implying that the maximal amount of electrons has been removed at this point and no further reduction is achieved.

This measurement proves that the magnetic pulse method removes stored electrons from the spectrometer volume and reduces the observed background from nuclear decays in the main spectrometer.

\begin{table}[h]
    \caption{Electron removal by the magnetic pulse with a \Kr{83m} source.
        The table shows the observed electron rate in the low-energy regime before and after a magnetic pulse, and the time $t$ of each pulse after start of the measurement.
        The removal efficiency $R$ in each pulse cycle was computed via \eqref{eq:removal_efficiceny}.}
    \newcolumntype{L}[1]{>{\raggedright\arraybackslash}p{#1}}
    \newcolumntype{C}[1]{>{\centering\arraybackslash}p{#1}}
    \newcolumntype{R}[1]{>{\raggedleft\arraybackslash}p{#1}}
    \newcommand{\ctab}{\centering\arraybackslash}
    \centering
    \begin{tabular}{rrrr}
    \toprule
         \ctab $t\slash\si{s}$
        &\ctab $\dot{N}_0\slash\si{cps}$
        &\ctab $\dot{N}_\mathrm{min}\slash\si{cps}$
        &\ctab $R$ \\
    \midrule
        \num{ 60}   &\num{2.47(4)}  &\num{1.52(4)}  &\num{0.38(2)}  \\
        \num{123}   &\num{1.37(2)}  &\num{1.02(2)}  &\num{0.25(2)}  \\
        \num{186}   &\num{1.22(2)}  &\num{0.87(2)}  &\num{0.29(2)}  \\
        \num{249}   &\num{1.13(2)}  &\num{0.81(3)}  &\num{0.28(3)}  \\
        \num{311}   &\num{1.12(2)}  &\num{0.79(2)}  &\num{0.30(3)}  \\
        \num{374}   &\num{1.07(4)}  &\num{0.76(2)}  &\num{0.29(3)}  \\
    \bottomrule
    \end{tabular}
    \label{tab:measurements:sds1_krypton}
\end{table}

\subsection{Phase II: Measurements with a radioactive \Rn{220} source}
\label{measurements:radon}

In the second commissioning phase of the spectrometer section the complete magnetic pulse system with flip-boxes was available to pulse all LFCS and EMCS air coils independently.
With this setup, the removal efficiency was again investigated with an artificially enhanced background; a \Rn{220}-emitting source ($t_{1/2} = \SI{55.6}{s}$) was attached to one of the spectrometer pump ports.
The \LN{}-cooled baffles at the pump ports are designed to prevent radon atoms from entering the spectrometer~\cite{Drexlin2017}; to achieve an increased background level in this setup the baffles were warmed up to about \SI{105}{K} to lower their blocking efficiency. 
The radon nuclei then decay in the spectrometer and produce low-energy stored electrons.
The observed rate reduction allows an investigation of the removal efficiency of the magnetic pulse under realistic conditions.

\begin{figure}[t]
    \centering
    \includegraphics[width=\columnwidth]{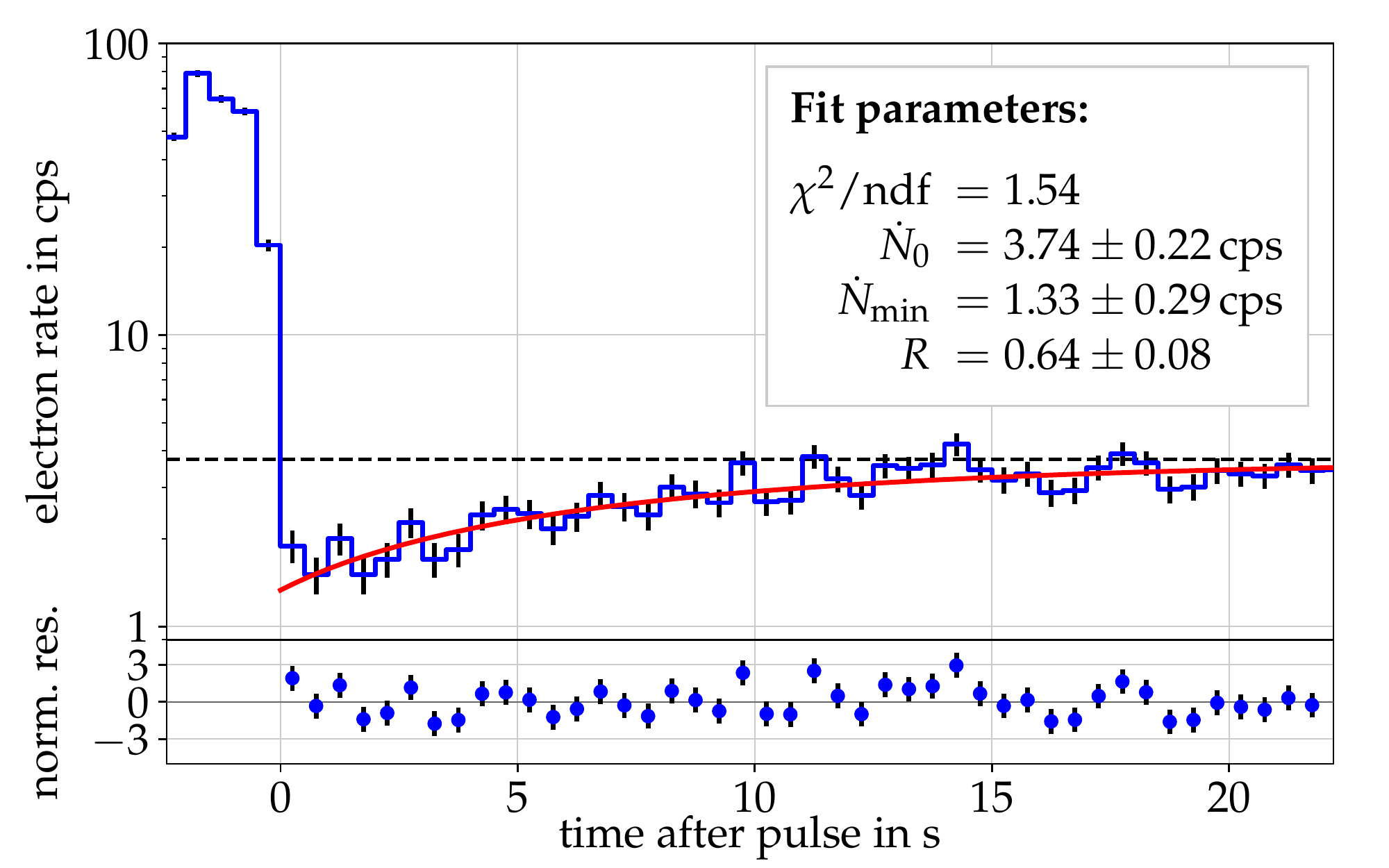}
    \caption{Removal of stored electrons originating from a \Rn{220} source by the magnetic pulse.
        The pulse was applied every \SI{25}{s} by inverting LFCS coils L1--L13 for \SI{1}{s} at a nominal magnetic field $B_\mathrm{min} = \SI{0.38}{mT}$.
        The plot shows the averaged electron rate from $n = \num{73}$ pulse cycles, which were aligned using the timing information from the reference signal.
        The electron rate between the pulse cycles is fit by an exponential model to determine the removal efficiency via \eqref{eq:removal_efficiceny}.
        The rate increases gradually during the relaxation period as the number of stored electrons increases from ongoing radon decays.}
    \label{fig:measurements:sds2_radon}
\end{figure}

A long-term measurement with the \Rn{220}-emitter and warm baffles was performed at the $B_\mathrm{min} = \SI{0.38}{mT}$ field setting.
Magnetic pulses with \SI{1}{s} pulse length were applied every \SI{25}{s}, inverting LFCS air coils L1--L13 simultaneously to produce a maximal field inversion.
Figure~\ref{fig:measurements:sds2_radon} shows the averaged electron rate of this long-term measurement with a total of \num{73} pulse cycles, using the ROI for low-energy secondary electrons from \tabref{tab:measurements:roi}.
This measurement was performed with an unbaked main spectrometer at a vacuum pressure of $\bigO{\SI{5e-10}{mbar}}$.
The start time of each pulse cycle is known with high accuracy from the reference trigger signal; therefore it is possible to align individual pulse cycles so that a summation can be performed to reduce the statistical uncertainty.
The figure shows the average rate of the combined \SI{0.5}{s} bins.
Note that the error bars (Poisson statistics, $\sigma = \sqrt{N}$) do not take the correlation of the radon-induced background events into account; for details see \cite{PhDHarms2015}.
The larger fluctuations of the radon-induced electron rate give rise to the rather high $\chi^2/\text{ndf}$ value of the fit.

During the relaxation period after the magnetic pulse, an exponential rate increase is observed due to continuous nuclear decays in the spectrometer volume.
By fitting the measurement data, we can determine the minimal electron rate $\dot{N}_\mathrm{min}$ at $t = 0$ (right after a magnetic pulse was applied) and the nominal background rate $\dot{N}_0$ for $t \rightarrow \infty$.
The fit result shows that the rate is reduced from an enhanced background rate of $\dot{N}_0 = \SI{3.74(22)}{cps}$ to $\dot{N}_\mathrm{min} = \SI{1.33(29)}{cps}$.
This yields a removal efficiency $R = \num{0.64(8)}$ as calculated via \eqref{eq:removal_efficiceny}.
The achieved background rate and the removal efficiency depend on the static magnetic field settings, which affect the electron storage conditions~\cite{PhDWandkowsky2013}.
A similar measurement at $B_\mathrm{min} = \SI{0.50}{mT}$ yields a comparable removal efficiency of $R = \num{0.57(26)}$.

The measurement results agree with the investigation discussed earlier, which used a \Kr{83m} source (\secref{measurements:krypton}).
It confirms the reduction of radon-induced background by the magnetic pulse under realistic conditions.
The difference between the removal efficiency of $R \approx \num{0.6}$ determined here and the result from the \Kr{83m} measurement, which yielded about half this value, can be attributed to the different setup.
The earlier measurement used only one flip-box instead of the fully-equipped LFCS, and also applied a different magnetic field setting.

\subsection{Phase II: Measurements at natural background level}
\label{measurements:natural}

The measurements with artificially increased background clearly show that the magnetic pulse method can reduce the background caused by nuclear decays inside the spectrometer volume.
Therefore we can now determine the removal efficiency without an artificial background source.
It is known that the \LN{}-cooled baffles in front of the pump ports efficiently block radon from the spectrometer.
Hence only a small amount of radon-induced background that can be targeted by active methods is expected.

\begin{figure}[t]
    \centering
    \includegraphics[width=\columnwidth]{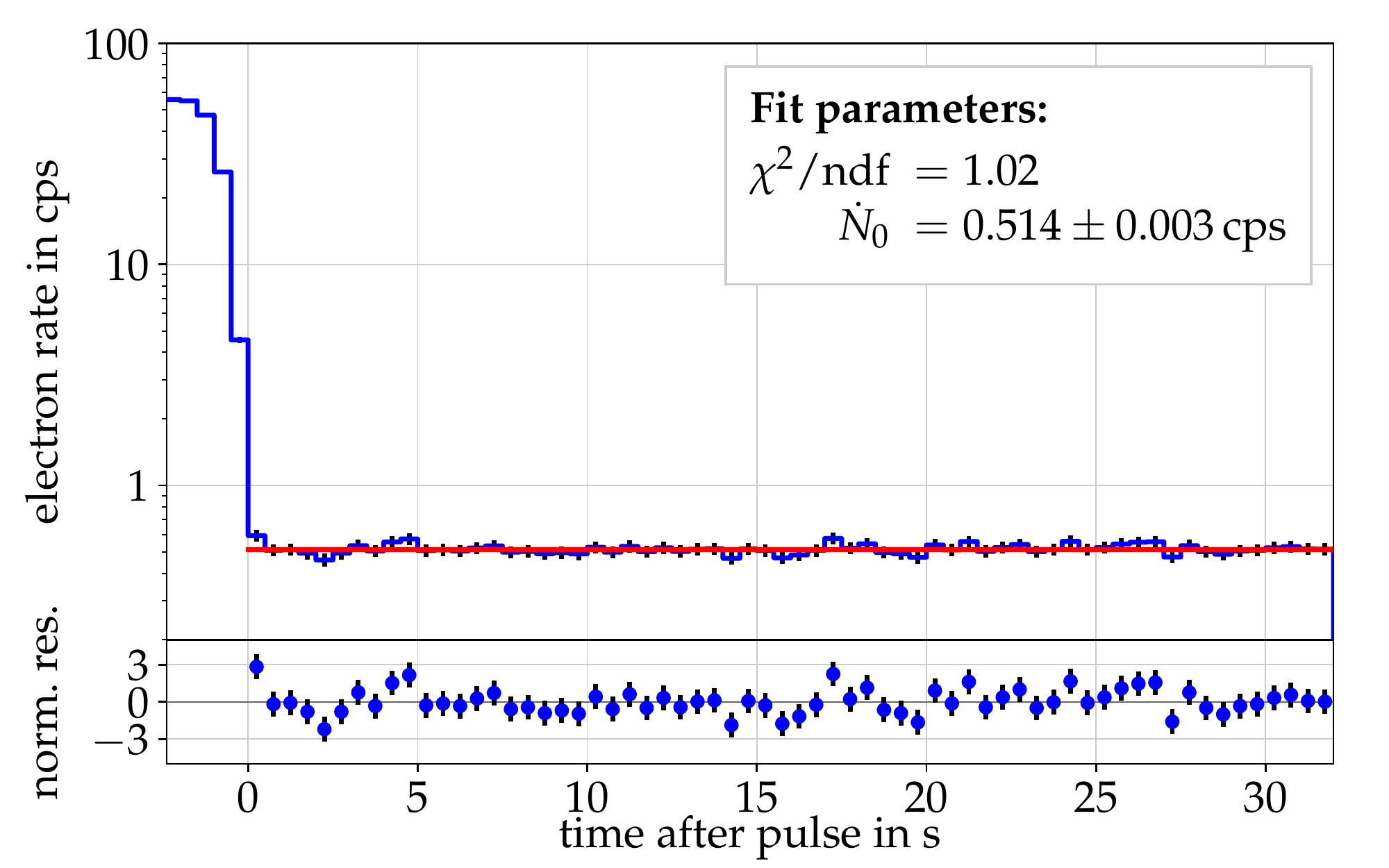}
    \caption{Effect of the magnetic pulse on the remaining background without artificial sources.
        The pulse was applied every \SI{35}{s} by inverting LFCS coils L1--L13 for \SI{1}{s} at a nominal magnetic field $B_\mathrm{min} = \SI{0.38}{mT}$.
        The plot shows the averaged electron rate from $n = \num{1544}$ pulse cycles aligned by the reference signal.
        The electron rate between the pulse cycles is fit by a linear model.
        In contrast to \figref{fig:measurements:sds2_radon} no background reduction is observed, although the visible ``pulse spike'' indicates that magnetic pulses are applied.}
    \label{fig:measurements:sds2_natural}
\end{figure}

Figure~\ref{fig:measurements:sds2_natural} shows the averaged electron rate in a long-term measurement using \LN{}-cooled baffles with \num{1544} pulse cycles over several hours, again using the secondary electron ROI from \tabref{tab:measurements:roi}.
In this case, a vacuum pressure of $\bigO{\SI{1e-10}{mbar}}$ was achieved with a baked spectrometer and activated getter pumps.
The pulses were applied with the same settings as in \secref{measurements:radon} at $B_\mathrm{min} = \SI{0.38}{mT}$.
The ``rate spike'' that was observed in other measurements is clearly visible, indicating that the magnetic pulse works as expected.
However, no rate reduction is observed and the measured electron rate is constant at $\dot{N}_0 = \SI{0.514(3)}{cps}$ over the pulse cycle.

Because it was shown earlier that the magnetic pulse method removes stored electrons from the spectrometer volume, this result indicates that the remaining background is \emph{not} caused by stored electrons that are typical of nuclear decays.
This confirms the efficiency of the \LN{}-baffles at blocking radon from the spectrometer volume~\cite{RadonPaper}.
Furthermore, this observation strongly indicates that electrons from the remaining background are presumably not stored (\ie electrons with $\transv{E} < \Delta E$, see \secref{design:background}).
The observation of a background level $> \SI{0.01}{cps}$ that cannot be reduced by the implemented passive and active methods provides further evidence of a novel background process at the main spectrometer.
This is in excellent agreement with investigations using the electric dipole method for background removal~\cite{PhDHilk2016}.
The background process would act on neutral particles that propagate from the vessel walls into the flux tube volume; a description is given in~\cite{Fraenkle2017}.

\subsection{Phase II: Measurements with an electron beam}
\label{measurements:egun}

A photo-electron source~\cite{Behrens2017} was installed at the main spectrometer entrance during commissioning measurements to investigate the transmission properties of the MAC-E filter~\cite{PhDBehrens2016,PhDErhard2016}.
The source produces a pulsed electron beam via the photo-electric effect, using an ultra-violet (UV) laser with a pulse frequency of \SI{100}{kHz} as a light source.
The emitted electrons have kinetic energies of $\sim \SI{18.6}{keV}$ and act as probes for the electromagnetic fields in the spectrometer.
Observing the disappearance of the electron beam at the detector allows the precise investigation of the timing characteristics of the magnetic field inversion.
When the magnetic field is inverted, the electrons are magnetically guided towards the vessel walls as per \eqref{eq:flux_tube}, instead of reaching the detector (see \figref{fig:magpulse_fluxtube}).
The time it takes for an electron at typical energies to travel through the spectrometer volume is short (a few \si{\micro{}s}) and can be neglected for this investigation.

Table~\ref{tab:measurements:egun} shows the pulse timing as measured with the electron beam for three positions on the pixelated detector wafer.
The three pixels correspond to different radial positions in the spectrometer, allowing the investigation of radial dependencies.
Again the air coils were operated at the $B_\mathrm{min} = \SI{0.38}{mT}$ setting.
At higher radii (pixels \#52 and \#100), the electron beam disappears earlier and reappears later than for the central position (pixel \#2).
This is expected due to the smaller distance between electron trajectory and vessel wall at higher radii.
In order to force the electron beam against the vessel walls, the magnetic field must be reduced sufficiently to shift the corresponding field line to a radius $\ge r_\mathrm{max}$.
The magnetic field $B_\mathrm{dis}$ where the electron beam disappears at the FPD can be estimated from \eqref{eq:flux_tube},
\begin{equation}
    \label{eq:beam_disappearance}
    B_\mathrm{dis} \approx \frac{\Phi(r_\mathrm{ana})}{\pi r_\mathrm{max}^2} \approx B_\mathrm{min} \cdot \frac{r_\mathrm{ana}^2}{r_\mathrm{max}^2}
    \,.
\end{equation}
Here $r_\mathrm{ana}$ is the radial position of a field line in the analyzing plane at magnetic field $B_\mathrm{min}$, and $\Phi(r_\mathrm{ana}) \le \SI{191}{T.cm^2}$ is the enclosed magnetic flux in the analyzing plane (see \figref{fig:magpulse_theory}).
The corresponding values are given in \tabref{tab:measurements:egun}.
As expected, electrons on central field lines require a considerably stronger field reduction to be removed, and beam disappearance is observed at a later time.
The disappearance times are consistent with the timing characteristics determined in \secref{measurements:test} ($\tau = \bigO{\SI{500}{ms}}$).
A comparison of the timing characteristics to simulation results is given in \secref{simulations:timing}.

\begin{table}[h]
    \caption{Timing of the magnetic pulse with a pulse length of \SI{2}{s} for different detector pixels.
        The table lists the corresponding field line radius in the analyzing plane $r_\mathrm{ana}$ and the beam disappearance/reappearance times $t_\mathrm{dis},\ t_\mathrm{re}$ during a pulse cycle.
        The required magnetic field $B_\mathrm{dis}$ to observe the electron beam disappearance was estimated from \eqref{eq:beam_disappearance}.}
    \newcommand{\ctab}{\centering\arraybackslash}
    \centering
    \begin{tabular}{rrrrr}
    \toprule
         \ctab $pixel \#$
        &\ctab $r_\mathrm{ana}\slash\si{m}$
        &\ctab $t_\mathrm{dis}\slash\si{sm}$
        &\ctab $t_\mathrm{re}\slash\si{ms}$
        &\ctab $B_\mathrm{dis}\slash\si{\micro T}$ \\
    \midrule
        \num{  2}   &\num{0.22}     &\num{409}      &\num{2275}     &\num{  0.8}    \\
        \num{ 52}   &\num{2.56}     &\num{308}      &\num{2371}     &\num{104.0}    \\
        \num{100}   &\num{3.45}     &\num{251}      &\num{2455}     &\num{188.3}    \\
    \bottomrule
    \end{tabular}
    \label{tab:measurements:egun}
\end{table}
%
\section{Simulations}
\label{simulations}

Particle-tracking simulations can provide insight to the electron storage conditions and the removal processes of the magnetic pulse.
For this we employed the simulation software \textsc{Kassiopeia} that has been developed over recent years by members of the KATRIN collaboration~\cite{Furse2017}.

\subsection{Implementation}
\label{simulations:setup}

The magnetic pulse was implemented in \textsc{Kassiopeia} as a new field computation module, which applies a time-dependent scaling factor $f(t)$ to a constant magnetic source field $\vec{B}_0(r)$.
The total magnetic field is given by the sum of the static contribution by the spectrometer solenoids and un-pulsed air coils (LFCS coil L14), and of the dynamic contribution by the pulsed air coils (LFCS coils L1--L13).

\begin{figure}[h]
    \centering
    \includegraphics[width=\columnwidth]{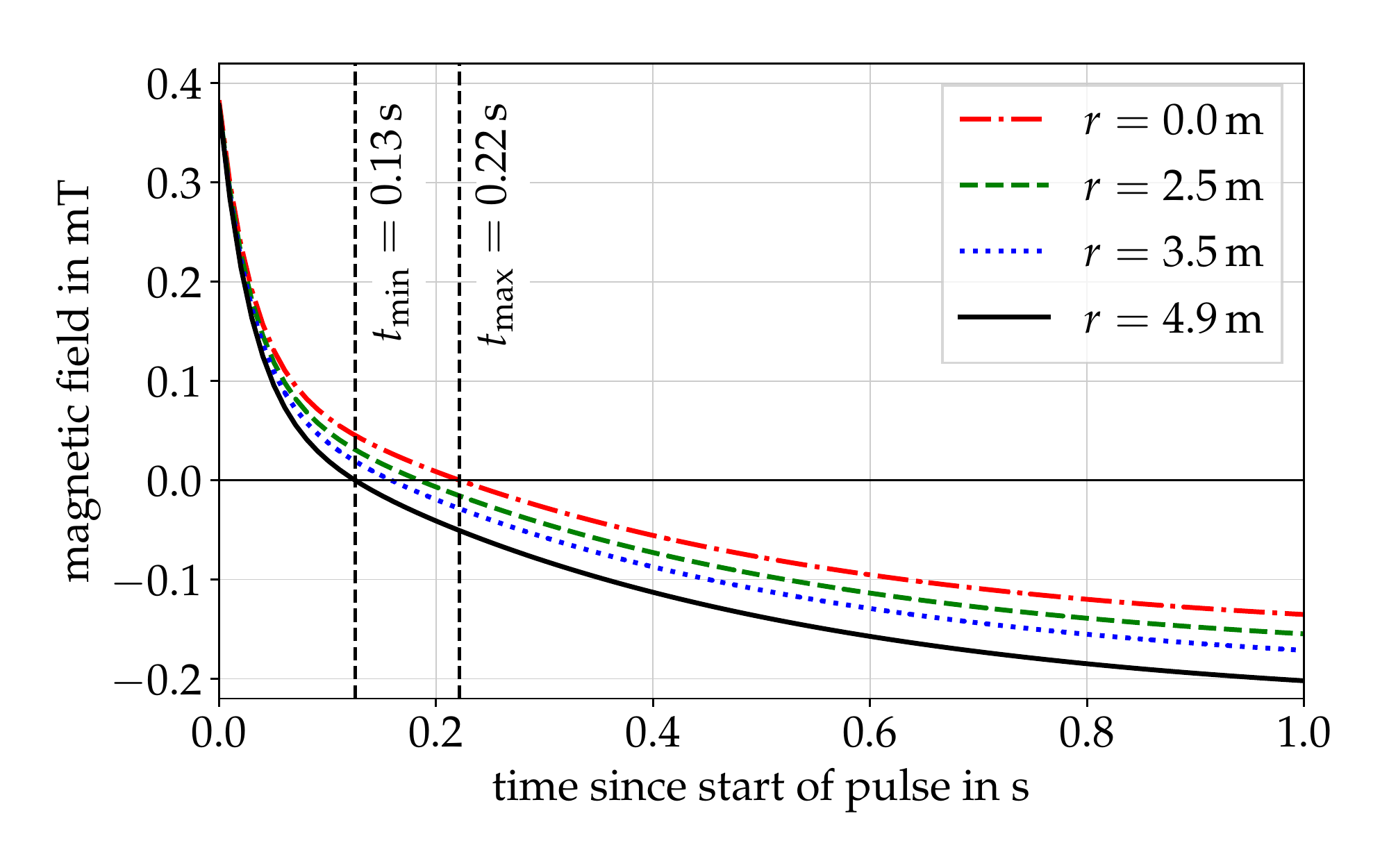}
    \caption{Simulated magnetic field during a magnetic pulse.
        The plot shows the value of $B_z(\vec{r}, t)$ from \eqref{eq:simulations:bfield} in the analyzing plane ($z = 0$) for different radii $r$.
        The dashed vertical lines indicate the minimal and maximal time $\hat{t}_\mathrm{min,max}$ when the field reaches zero.
        Due to the radial inhomogeneity of the magnetic field, this time is shorter for higher radii and minimal at the vessel radius $r_\mathrm{max} = \SI{4.9}{m}$.}
    \label{fig:simulations:magpulse_timing}
\end{figure}

The simulations presented here use a double-exponential time-dependency that matches field measurements at the spectrometer (\secref{measurements:test}):
\begin{align}
    \label{eq:simulations:bfield}
    \vec{B}(\vec{r}, t) &= f(t) \cdot \vec{B}_0(\vec{r}) + \vec{B}_\mathrm{static}(\vec{r})
    \,,
\\
    f(t) &= \exp\left( \frac{-t}{\tau_1} \right) + \exp\left( \frac{-t}{\tau_2} \right) - 1
    \,,
\end{align}
where $\tau_{1} = \SI{30}{ms}$ and $\tau_2 = \SI{420}{ms}$ are the time constants used in the simulation, $\vec{B}_0(\vec{r})$ is the source field, and $\vec{B}_\mathrm{static}(\vec{r})$ the static (unmodified) field at the electron's position $\vec{r}$.
Figure~\ref{fig:simulations:magpulse_timing} illustrates the time-dependence of the simulated magnetic field.
The long-term behavior of the magnetic field change is the relevant timescale for electron removal, therefore the outcome of the simulation is not strongly dependent on accurate field modeling on short timescales\footnote{%
    For $t < \SI{100}{ms}$ the time constant cannot be determined exactly. Deviations from the exponential behavior in \eqref{eq:simulations:bfield} are attributed to the mutual inductance of the individual air coils and eddy currents in the spectrometer vessel walls. See \cite{Erhard2018} for details.}.

The induced electric field that results from the magnetic field change must be taken into account as well.
The dominating azimuthal field component $E_\phi$ is superimposed on the overall electric field by an additional field module.
According to \eqref{eq:magpulse_induced_field} it is defined by the derivative of the scaling factor,
\begin{align}
    \label{eq:simulations:efield}
    E_\phi(\vec{r}, t) &= -\frac{r}{2} \cdot \dot{B}_z(\vec{r}, t)
                        = -\frac{r}{2} \cdot f'(t) \cdot B_{0,z}(\vec{r})
    \,,
\\
    f'(t) &= -\frac{1}{\tau_1} \cdot \exp\left( \frac{-t}{\tau_1} \right)
             -\frac{1}{\tau_2} \cdot \exp\left( \frac{-t}{\tau_2} \right)
    \,,
\end{align}
using the same variables as before and with $r$ the radial position of the electron in the spectrometer.

For the Monte-Carlo (MC) simulations we employed a quasi-static approach to reduce the computation time.
For each time-step in an interval $t_\mathrm{S} = \SIinterval{0}{1}{s}$ with \SI{10}{ms} step size, a MC simulation is carried out with a fixed magnetic field $\vec{B}(\vec{r}, t = t_\mathrm{S})$.
The first step at $t_\mathrm{S} = 0$ corresponds to nominal magnetic field at the start of a magnetic pulse cycle, and subsequent time-steps allow an investigation of the storage conditions over time.
This approach is justified since the particle-tracking times in the simulation are short ($\ll \SI{1}{ms}$) compared with the magnetic field change ($\tau > \SI{100}{ms}$), so that the magnetic field can be assumed constant at each step.
Because the electron storage conditions are mainly defined by the magnetic field, the simulations used a simplified setup for the electrode geometry which only included the spectrometer vessel at $U = \SI{-18.4}{kV}$.

\subsection{Removal efficiency}
\label{simulations:efficieny}

\begin{figure*}[t]
    \centering
    \includegraphics[width=\columnwidth]{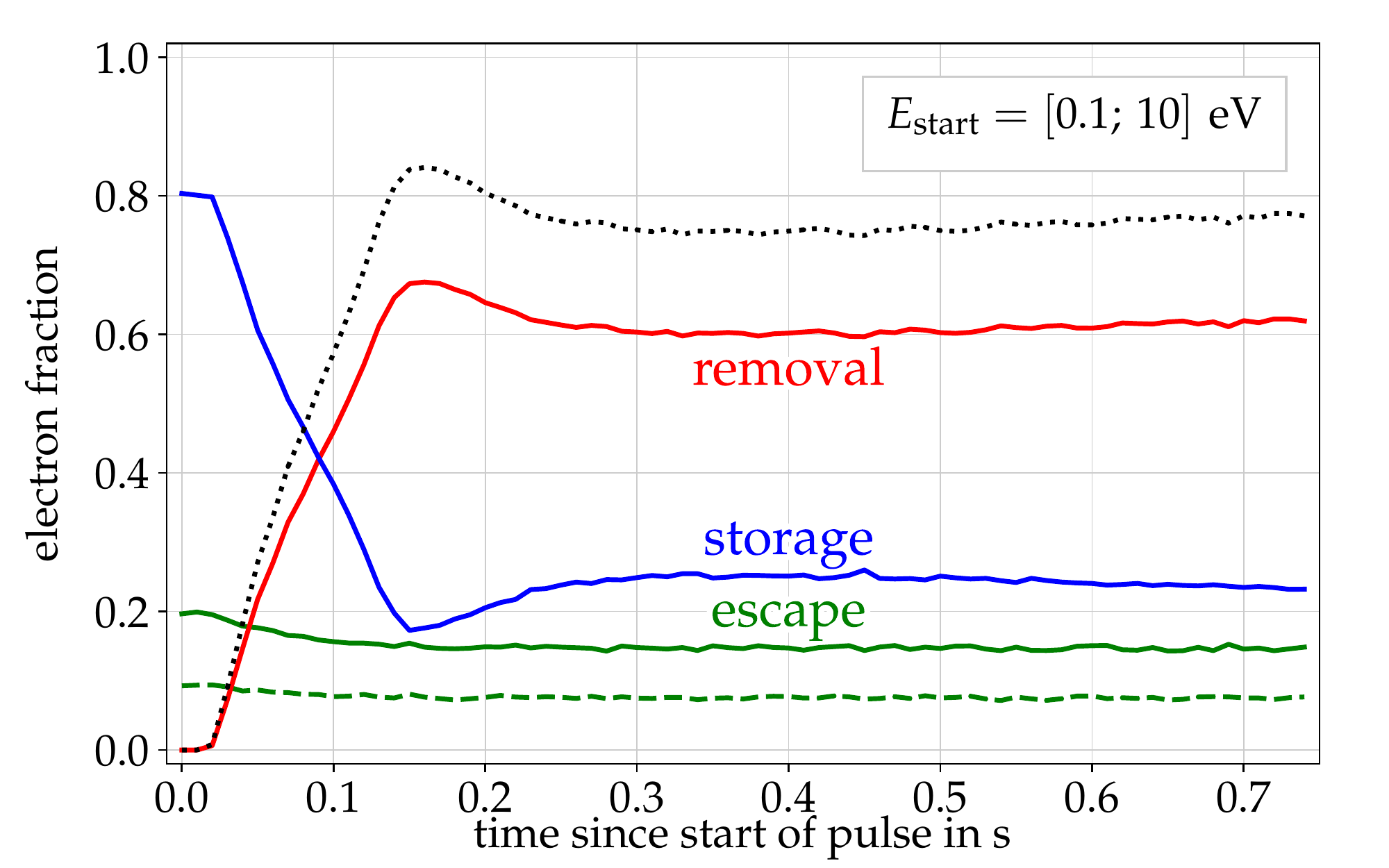}
    \includegraphics[width=\columnwidth]{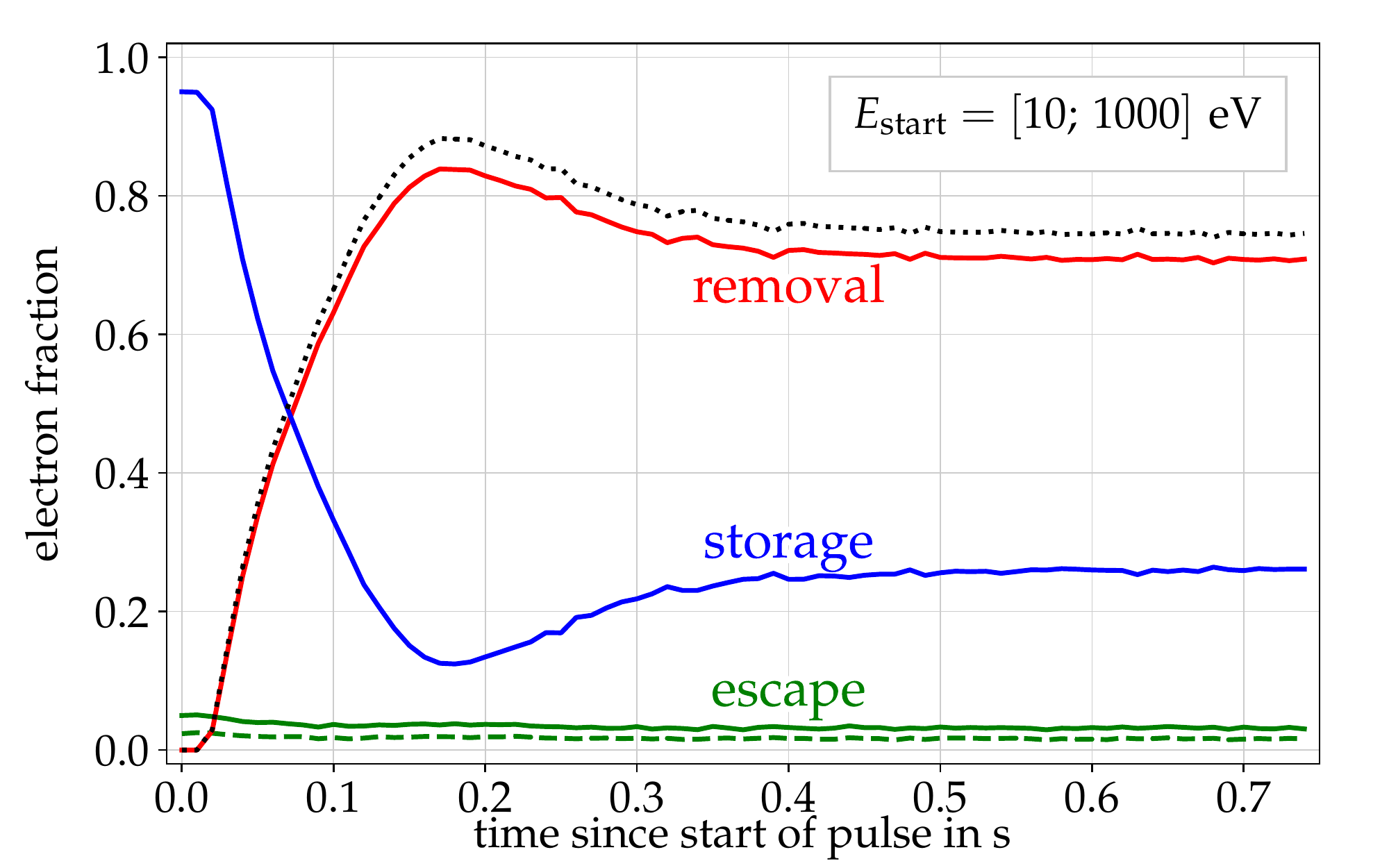}
    \caption{Simulated removal efficiency of the magnetic pulse in different energy regimes.
        The plots show the fraction of removed, stored, and escaping electrons (solid lines) from a total amount of \num{25000} electrons created randomly inside the nominal flux tube volume.
        The dotted line shows the removal efficiency from \eqref{eq:simulated_removal_efficiceny}.
        \emph{Left:} Low-energy electrons have a storage probability of about 80\% under nominal conditions.
        \emph{Right:} High-energy electrons are stored more efficiently with a probability of about 95\%.
        In both cases, about one half of the escaping electrons reach the detector (dashed lines at the bottom); these electrons are observable in measurements.
        The magnetic pulse achieves a transfer from stored to removed electrons and the removal efficiency reaches a maximum at $t \approx \SI{0.16}{s}$ in both energy regimes.}
    \label{fig:simulations:removal_efficiency}
\end{figure*}

The measurements at the main spectrometer (\secref{measurements:radon}) showed a reduction of the stored-electron induced background with a removal efficiency of $R \approx \num{0.6}$, which indicates that stored electrons are not entirely removed by the magnetic pulse.
Simulations are used to investigate the removal efficiency in detail.
The reduced magnetic field affects stored electrons due to several processes that were explained in \secref{design:bgreduction}.
The fraction of stored electrons that can be removed is determined from MC simulations, where a large population of electrons is generated in the spectrometer volume at starting time $t_0 = t_\mathrm{S}$ with random initial energy $E_0$ (uniform distribution) and pitch angle $\theta_0$ (isotropic distribution).
The initial energy distribution was split up into a low-energy regime with $E_0 = \SIinterval{0.1}{10}{eV}$ and a high-energy regime with $E_0 = \SIinterval{10}{1000}{eV}$.
A more detailed investigation that also covers the energy regime up to \SI{100}{keV} is available in \cite{PhDBehrens2016}.

In the simulation, these electrons were tracked until one of three termination conditions was met:
\begin{itemize}
    \item
        The electron exits through the spectrometer entrance or exit (axial position $|z| \ge \SI{+-12.2}{m}$).
        This electron \emph{escapes} from the storage volume in direction of the source or detector.

    \item
        The electron hits the inner surface of the spectrometer vessel (radial position $r \ge r_\mathrm{max}(z)$, where $r_\mathrm{max}(z) \le \SI{4.9}{m}$ is the spectrometer vessel radius at axial position $z$) and is considered to be \emph{removed} from the spectrometer.
        At nominal magnetic field, the flux tube is fully contained inside the spectrometer, therefore this effect is only observed when the magnetic field is sufficiently reduced.

    \item
        The electron is reflected twice inside the spectrometer volume.
        A reflection is indicated by the condition $(\vec{B}\cdot\vec{p}) \, (\vec{B}'\cdot\vec{p}') < 0$, which is equivalent to a change of direction along the electron trajectory. $\vec{B}$ indicates the magnetic field and $\vec{p}$ the electron momentum; the dashed symbols denote the values from the previous step in the simulation.
        In this case the electron is considered to be \emph{stored} since it does not escape the spectrometer volume.

        Note that electrons could also be removed from the flux tube by non-adiabatic propagation, which results in a ``chaotic'' trajectory and increases the electron's chance to hit the vessel walls.
        However, in the energy range and magnetic field setting considered here ($E < \SI{1}{keV}$, $B_\mathrm{min} \approx \SI{0.3}{mT}$) these effects do not play a significant role.
\end{itemize}

Using the approach discussed in \secref{simulations:setup}, varying the time $t_\mathrm{S}$ allows the examination of the storage probability and the removal efficiency over a complete pulse cycle.
A total of $N = \num{25000}$ electrons were started in each \SI{10}{ms} bin of the starting time $t_0 = t_\mathrm{S} = \SIinterval{0}{1}{s}$ in the simulation.

Here the removal efficiency is defined as the ratio between the fraction of removed electrons at the time $t$ after a magnetic pulse and the fraction of stored electrons at nominal conditions,
\begin{equation}
    \label{eq:simulated_removal_efficiceny}
    R_\mathrm{sim}(t) = \left[\frac{N_\mathrm{removed}}{N}\right]_{t_\mathrm{S} = t} \;\cdot\; \left[\frac{N_\mathrm{stored}}{N}\right]^{-1}_{t_\mathrm{S} = 0}
    \;\,.
\end{equation}
The value $R(t)$ therefore puts the amount of electrons that is removed at a given time $t$ during a magnetic pulse ($N_\mathrm{removed}$) into relation with the amount of electrons that is stored under nominal conditions at $t_\mathrm{S} = 0$ ($N_\mathrm{stored}$).
This definition has the advantage that it does not depend on a constant number of simulated electrons\footnote{%
    In some simulation runs the number of electron tracks is smaller than \num{25000} due to occasional numerical errors in the trajectory calculation; these tracks are excluded from the analysis.}.
Electrons which escape the spectrometer are not considered here since their fraction is not significantly affected by the magnetic pulse (see \figref{fig:simulations:removal_efficiency}).

Equation~\ref{eq:simulated_removal_efficiceny} provides a good estimate of the net-effect of the magnetic pulse method, which can be compared to measurement results discussed in \secref{measurements}.
One expects the value $R_\mathrm{sim}(t)$ to increase over a magnetic pulse cycle while the magnetic field is reduced.
The removal efficiency of a complete cycle corresponds to the maximum value of $R_\mathrm{sim}(t)$ during the given time frame, which is denoted here as $\hat{R}_\mathrm{sim}$.
Because electrons once removed do not re-enter the flux tube on a short timescale, the maximum value describes the largest possible fraction of electrons that is removed by a single pulse.

Figure~\ref{fig:simulations:removal_efficiency} shows the fraction of escaping, removed, and stored electrons and the corresponding removal efficiency from \eqref{eq:simulated_removal_efficiceny} during a magnetic pulse cycle.
Under nominal conditions, the majority of electrons are stored in both energy regimes; the storage probability is \SI{80}{\%} for low-energy and \SI{95}{\%} for high-energy electrons.
The remaining electrons escape from the spectrometer, with about \SI{50}{\%} of these electrons reaching the detector.
At $t = \SI{0.16}{s}$, the removal efficiency \eqref{eq:simulated_removal_efficiceny} reaches a maximum of $\hat{R}_\mathrm{sim} = \numrange{0.84}{0.88}$.
The maximum efficiency is reached at nearly the same time in both energy regimes; it corresponds to a maximum reduction of the absolute magnetic field in the outer region of the central spectrometer section (\figref{fig:simulations:magpulse_timing}).

The removal efficiency $\hat{R}_\mathrm{sim}$ is considerably higher than the measurement result in \secref{measurements:radon}, which yielded a value of \num{0.64(8)}.
One possible explanation for this discrepancy is that the fraction of stored electrons is estimated incorrectly by the simulation.
The termination conditions discussed above consider an electron stored if it was reflected twice.
However, in principle these electrons could escape by non-adiabatic propagation after multiple reflections later in the simulation; in this case $N_\mathrm{stored}$ determined from the simulation would be somewhat smaller in reality.
This would overestimate $\hat{R}_\mathrm{sim}$ and could explain a discrepancy of a few percent to the measurement result.
The energy dependency of the removal efficiency could also play a role; however, \figref{fig:simulations:removal_efficiency} indicates that $R_\mathrm{sim}(t)$ has only little dependency on the electron energy.
The simulation also applies a quasi-static approach and therefore neglects dynamic processes that could play a role in electron removal, and further assumes a time-dependency of the magnetic field that is based on measurements outside of the vessel.
It therefore does not include modifications due to eddy currents.
A different time-dependency could contribute to the observed deviation.
Nevertheless, the simulation shows a good qualitative agreement to the measurement with \Rn{220}, and thus allows an investigation of the overall behavior and the relevant removal processes.

For $t \ge \hat{t}$ the storage conditions are broken by the reduced magnetic field, and the majority of electrons are removed at the vessel walls.
In this time interval, the removal efficiency diminishes because some removal processes depend on the absolute value $|B|$ of the magnetic field, so that the overall efficiency decreases as $B$ becomes more negative (see \figref{fig:simulations:magpulse_timing}).
The observation that the removal efficiency does not reach \SI{100}{\%} is explained by new electron storage conditions that are created in the central spectrometer volume; this effect is further investigated in \secref{simulations:storage}.
At $t > \SI{1}{s}$  (not shown in the figure), the magnetic field returns to nominal strength.

Figure~\ref{fig:simulations:removal_efficiency} also shows that the fraction of escaping electrons remains comparable to nominal conditions for $t > 0$.
For the low-energy regime, the fraction of escaping electrons reduces from \num{0.2} at nominal conditions to \num{0.15} at $t \approx \SI{0.75}{s}$.
This is explained by the typically small pitch angles of the escaping electrons, which results in a low storage probability that does not strongly depend on the magnetic field.

\subsection{Pulse timing}
\label{simulations:timing}

The simulations also allow an examination of the timing of the magnetic field inversion.
In the commissioning measurements at the main spectrometer, a pulsed electron beam was used to observe the beam disappearance at the FPD (\secref{measurements:egun}), which corresponds to the time when the magnetic field lines connect to the vessel walls.
With simulations it is possible to determine the time when a field line, corresponding to a specific detector pixel, touches the vessel walls.

The simulated disappearance times for a typical magnetic pulse were determined as follows.
Field lines are tracked from different detector pixels at starting times $t_0 = t_\mathrm{S} = \SIinterval{0}{0.5}{s}$ with a step size of \SI{1}{ms}; for each starting time the magnetic field is scaled according to \eqref{eq:simulations:bfield}.
The time when a field line connects to the vessel walls corresponds to the disappearance time $t_\mathrm{dis,sim}$ with respect to the start of the pulse cycle at $t_\mathrm{S} = 0$.

The results are compared to the measurements in \tabref{tab:simulations:egun}; measurement data are only available for three detector pixels.
An average discrepancy of $\overline{\Delta t_\mathrm{dis}} = \SI{247}{ms}$ for these pixels is observed.
The discrepancy shows no clear dependency on the detector pixel and indicates that the overall magnetic field reduction is delayed in comparison with the simulation.
In the measurement, the start time of the magnetic pulse is known precisely from the reference trigger signal, therefore the delay must be explained by physical effects that slow down the magnetic field change.
One natural explanation is eddy currents in the stainless steel hull of the spectrometer vessel.
In addition, the air coils behave as a coupled system due to the small distance between adjacent air coils, which is small compared with the coil radius~\cite{Erhard2018}.
Hence, the mutual inductance plays a significant role that can further slow down the magnetic field change.
The observed delay is attributed to a combination of these effects.
However, because the magnetic pulse is typically applied with durations of \SI{500}{ms} or more, this the delay does not affect its removal efficiency in practice.

\begin{table}[htb]
    \centering
    \caption{Pulse timing for different detector pixels. The table shows the measured and simulated beam disappearance times at the detector, $t_\mathrm{dis}$ and $t_\mathrm{dis,sim}$, and the calculated difference $\Delta t_\mathrm{dis} = t_\mathrm{dis} - t_\mathrm{dis,sim}$. The individual detector pixels correspond to different radial positions $r_\mathrm{ana}$ in the analyzing plane of the spectrometer.}
    \newcommand{\ctab}{\centering\arraybackslash}
    \begin{tabular}{rrrrrr}
    \toprule
         \ctab $pixel \#$
        &\ctab $r_\mathrm{ana}\slash\si{m}$
        &\ctab $t_\mathrm{dis,sim}\slash\si{ms}$ 
        &\ctab $t_\mathrm{dis}\slash\si{ms}$
        &\ctab $\Delta t_\mathrm{dis}\slash\si{ms}$ \\
    \midrule
        \num{  2}   & \num{0.22}    &\num{153}      &\num{409}      &\num{256}      \\
        \num{  4}   & \num{1.06}    &\num{102}      &---            &---            \\
        \num{ 28}   & \num{1.97}    &\num{ 67}      &---            &---            \\
        \num{ 52}   & \num{2.56}    &\num{ 48}      &\num{308}      &\num{260}      \\
        \num{ 75}   & \num{2.79}    &\num{ 41}      &---            &---            \\
        \num{100}   & \num{3.45}    &\num{ 25}      &\num{251}      &\num{226}      \\
        \num{124}   & \num{3.82}    &\num{ 18}      &---            &---            \\
    \bottomrule
    \end{tabular}
    \label{tab:simulations:egun}
\end{table}

\subsection{Electron storage conditions}
\label{simulations:storage}

\begin{figure*}[t]
    \centering
    \includegraphics[width=\columnwidth]{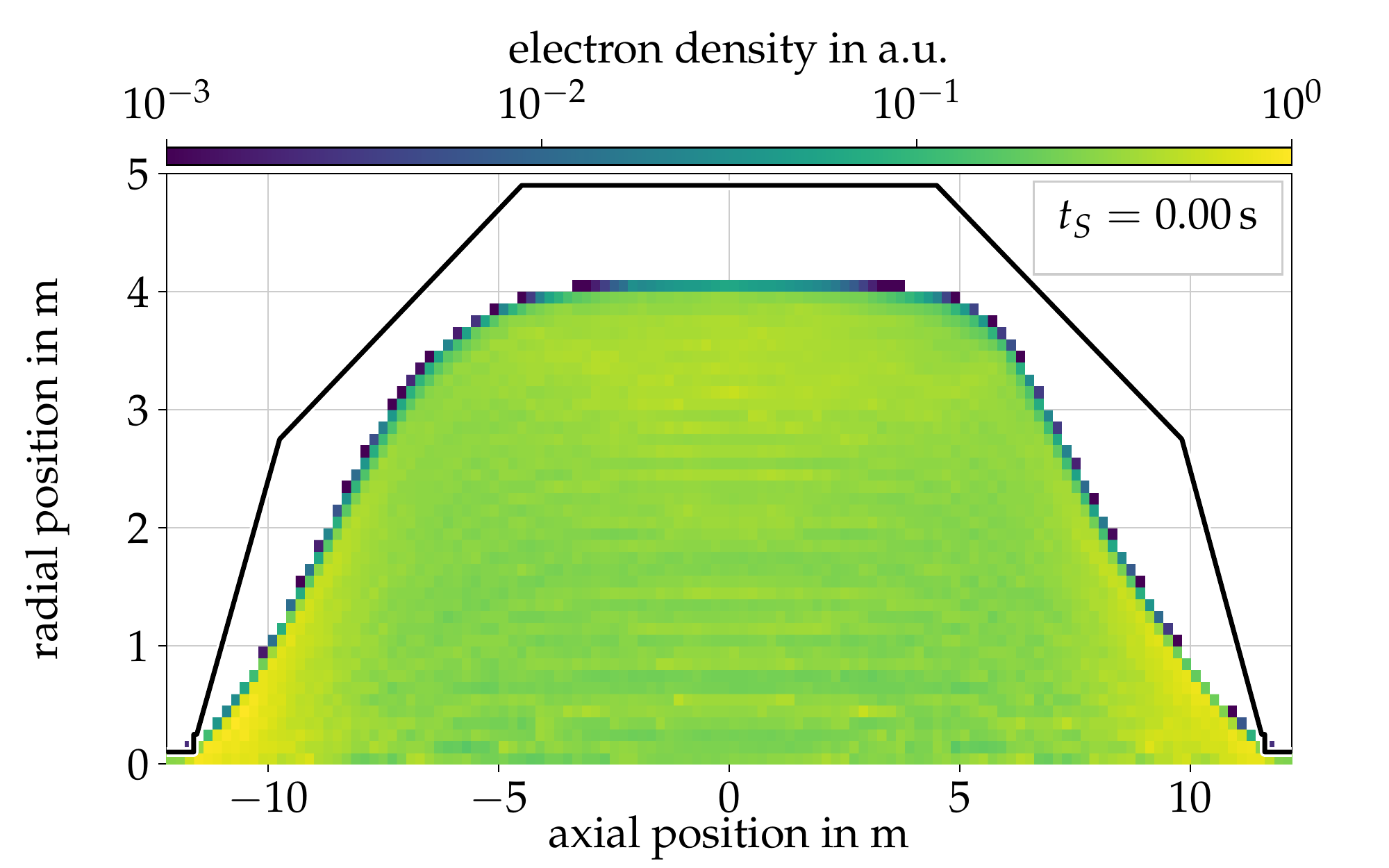}
    \includegraphics[width=\columnwidth]{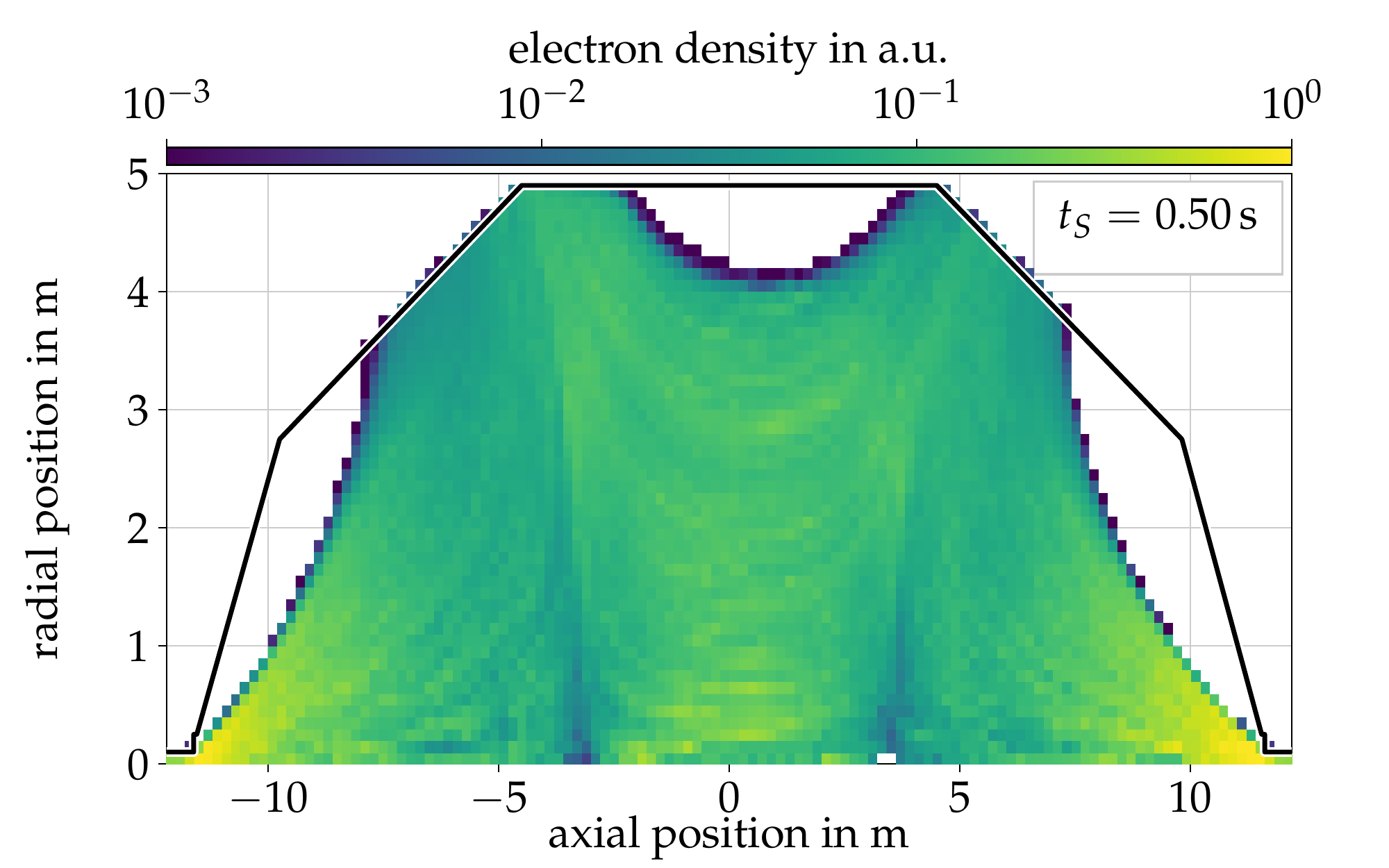}
    \caption{Simulated electron density in the flux tube during a magnetic pulse.
        The plots show the density rotated around the spectrometer axis (axial symmetry).
        The black outline indicates the spectrometer walls with $r_\mathrm{max} = \SI{4.9}{m}$.
        The electron density was determined by filling each step of the simulated electron trajectories into a two-dimensional $(r,z)$-histogram and weighting by \eqref{eq:electron_density_simulated}.
        \emph{Left:} The normalized density of low-energy electrons with $E_0 = \SIinterval{0.1}{10}{eV}$ at nominal conditions ($t_\mathrm{S} = \SI{0}{s}$) is almost homogeneous in the flux tube volume.
        \emph{Right:} At inverted magnetic field ($t_\mathrm{S} = \SI{0.5}{s}$), the flux tube is strongly deformed and electrons are removed from the nominal flux tube, resulting in regions with reduced density.
        A region around the spectrometer center ($|z| \lesssim \SI{5}{m}$) remains where electrons are stored despite the inverted field.}
    \label{fig:simulations:density_map}
\end{figure*}

A deeper understanding of the magnetic pulse can be gained by examining the electron density in the spectrometer during a pulse cycle.
The electron density can be computed directly from the simulation results discussed in \secref{simulations:efficieny}.
Here the density is determined by filling each step $i$ of a simulated electron trajectory into an $(r, z)$-histogram with bin size $\Delta r = \SI{0.1}{m}$ and $\Delta z = \SI{0.2}{m}$.
Each step is weighted by the time $\Delta t_i$ the electron spends in one bin, which corresponds to a cylindrical shell in the spectrometer volume. The weight $w_i$ is then normalized to the bin volume to compare the density at different radii:
\begin{equation}
    \label{eq:electron_density_simulated}
    w_i(r, z) = \frac{\Delta t_i(r, z)}{\pi \left((r + \Delta r)^2 - r^2\right) \cdot \Delta z}
    \,.
\end{equation}
The denominator corresponds to the volume of a bin with dimension $(\Delta r, \Delta z)$, and the numerator is the time spent in the bin $(r, z)$.
The time must be considered here to correctly take electrons with different kinetic energies into account.
The simulations thus allow the investigation of the spatial distribution of the electron storage conditions in the spectrometer volume and their time-dependency.

Figure~\ref{fig:simulations:density_map} shows two electron density maps that correspond to the conditions at $t_\mathrm{S} = 0$ (nominal magnetic field, $B_\mathrm{min} = \SI{0.38}{mT}$) and $t_\mathrm{S} = \SI{0.5}{s}$ (inverted magnetic field; see \figref{fig:simulations:magpulse_timing}).
The density maps are shown here only for the low-energy regime with $E_0 = \SIinterval{0.1}{10}{eV}$; high-energy electrons show a similar behavior.
The electron density in the figure is given in arbitrary units and rescaled to a relative range \numrange{e-3}{1} to allow a qualitative comparison between the different electromagnetic conditions.
Details are given in \cite{PhDBehrens2016}.

At $t_\mathrm{S} = 0$ (left panel), the electron density is nearly constant over the entire flux tube.
An increased density is observed at the entrance and exit regions of the spectrometer, where electrons are confined to a smaller volume.

When the magnetic field is reduced at $t_\mathrm{S} > 0$ (right panel), the flux tube widens and the outer parts of the flux tube volume touch the vessel walls.
Electrons that were stored in the outer flux tube under nominal conditions are now removed.
After the magnetic field is inverted, its magnitude $|B|$ increases while more field lines connect to the vessel walls.
It now becomes possible for electrons to be magnetically reflected while propagating along a field line, which prevents them from being removed at the vessel walls.
From the simulation data one can easily determine the volume in which electrons become trapped (see definition in \secref{simulations:efficieny}).
It follows that the storage region is confined to $|z| \lesssim \SI{5}{m}$ in the examined setting; its extent features a considerable radial dependency as visualized in the right panel of \figref{fig:simulations:density_map}.
Because magnetic reflection results from a transformation of the pitch angle $\theta$ in an inhomogeneous magnetic field according to \eqref{eq:magnetic_moment}, this affects mainly electrons with large initial pitch angles.
Unfortunately, these are the electrons that are stored most efficiently under nominal conditions for the same reason.

With the current setup of the magnetic pulse system that is based on inverting air coil currents, it is impossible to circumvent the magnetic bottle effect that arises during a magnetic pulse cycle.
The remaining electrons stored in the central spectrometer volume therefore cannot be removed by the magnetic pulse alone, and the total removal efficiency of this method is limited.
This agrees with measurements that indicated a strong, but less-than-maximal background reduction by the magnetic pulse method (\secref{measurements:radon}), and the corresponding simulations (\secref{simulations:efficieny}).
%
\section{Conclusion}
\label{conclusion}

In this work we presented the theory, design, and comissioning of a novel background reduction technique at the KATRIN experiment, the so-called  magnetic pulse method.
Our implementation inverts the currents of the individual air coils that surround the main spectrometer, which achieves a reduction or inversion of the magnetic guiding field on short timescales.
Dedicated current-inverter units (``flip-boxes'') were designed for this purpose, they handle air coil currents up to their maximum design value of \SI{175}{A}.
In addition to enabling the removal of stored electrons by a magnetic pulse, the flip-box setup greatly enhances the flexibility of the existing air coil system.
It enables measurements with special magnetic field settings in which selected air coils are operated at inverted current, which allows a variety of dedicated background measurements.

We discussed measurements at the KATRIN main spectrometer with a preliminary system that used a single flip-box prototype, and with the fully implemented system that consists of \num{16} flip-boxes.
Measurements were performed with radioactive sources (\Kr{83m} and \Rn{220}) to artificially increase the background from nuclear decays, and with nominal spectrometer background where radon decays in the spectrometer volume are efficiently suppressed.
These measurements clearly show that the magnetic pulse method can remove stored electrons from the magnetic flux tube.
At the nominal magnetic field setting, the determined removal efficiency of a single magnetic pulse is about \num{0.6} for low-energy secondary electrons that originate from \Rn{220} decays.
The method is therefore suitable to suppress spectrometer background that is induced by nuclear decays of radioactive isotopes such as radon.
Our measurements at nominal background (with all \LN{}-baffles cold) showed no reduction of the observed electron rate.
This is attributed to the highly efficient suppression of background from stored electrons by the passive background reduction methods implemented at the main spectrometer.

Like the magnetic pulse, the complementary active background reduction method that applies an electric dipole field targets electrons that are stored in the spectrometer volume.
One would therefore not expect a large improvement in background reduction by combining both methods.
However, the inefficiency of both methods in removing the remaining background strongly implies that this background is not caused by stored electrons. 
Instead, it is more likely that neutral messenger particles that enter the magnetic flux tube create low-energy background electrons; this background cannot be removed by the electric dipole or the magnetic pulse.

In addition to measurements, we examined the removal processes in more detail by particle-tracking simulations with the \textsc{Kassiopeia} software.
We found that with the implementation of the magnetic pulse method described in this article, the removal efficiency is intrinsically limited due to the complex electromagnetic conditions in the main spectrometer volume.
The inversion of the magnetic guiding field, which is accompanied by a considerable deformation of the magnetic flux tube, creates new electron storage conditions in the central spectrometer region.
This prevents a complete removal of stored electrons from the flux tube, and a fraction of stored electrons remains after a magnetic pulse cycle.
A possible improvement could be to adapt the design and change how the magnetic field reduction is applied through the air-coil system.

The magnetic pulse method provides an efficient technique to remove stored electrons from the main spectrometer flux tube, and is a viable enhancement of the existing large-volume air coil system.
The method targets stored electrons with a wide range of kinetic energies, including high-energy primary electrons from nuclear decays.
Although the active background removal techniques currently cannot significantly reduce the observed spectrometer background, they may contribute to a background reduction in future measurement phases where we expect a lower overall background level.
Furthermore, the active methods are not specifically targeting background from radon decays and therefore provide a suitable technique to remove stored electrons that originate from other sources.
\begin{acknowledgements}
We acknowledge the support of Helmholtz Association (HGF), Ministry for Education and Research BMBF (05A17PM3, 05A17PX3, 05A17VK2, and 05A17WO3), Helmholtz Alliance for Astroparticle Physics (HAP), and Helmholtz Young Investigator Group (VH-NG-1055) in Germany; Ministry of Education, Youth and Sport (CANAM-LM2011019), cooperation with the JINR Dubna (3+3 grants) 2017--2019 in the Czech Republic; and the Department of Energy through grants DE-FG02-97ER41020, DE-FG02-94ER40818, DE-SC0004036, DE-FG02-97ER41033, DE-FG02-97ER41041, DE-AC02-05CH11231, and DE-SC0011091 in the United States.
\end{acknowledgements}
%
\bibliography{RefMagpulse,references/RefArXivPublications,references/RefGeneralPublications,references/RefKATRINDiplomaMasterTheses,references/RefKATRINPhDTheses,references/RefKATRINPublications}
\bibliographystyle{spphys-mod}
%
%
\vfill
\end{document}